\pdfoutput=1

\documentclass[11pt,a4paper]{article}
\usepackage{bm}
\usepackage{jheppub}
\usepackage{bbm}
\usepackage{mathrsfs}

\allowdisplaybreaks

\let\de=\partial
\let\eps=\epsilon
\let\k=\kappa
\let\l=\lambda
\let\m=\mu
\let\n=\nu
\let\a=\alpha
\let\b=\beta
\let\t=\theta
\let\x=\xi
\let\om=\omega

\newcommand\im{\text{i}}
\newcommand\gr[1]{\mathrm{#1}}
\newcommand\Aa{\mathscr{A}}
\newcommand\La{\mathscr{L}}
\newcommand\Pe{\mathscr{P}}
\newcommand\dd{\text{d}}
\newcommand\gal{\mathfrak{g}}
\newcommand\hh{\Xi}
\newcommand\str{\Lambda}
\newcommand\Ka{t}
\DeclareMathOperator{\sgn}{sgn}
\newcommand\LL{\amalg}
\newcommand\XX{\Pi}
\DeclareMathOperator{\Co}{\widehat{ch}}
\DeclareMathOperator{\Si}{\widehat{sh}}

\newtheorem{theorem}{Theorem}
\definecolor{myred}{rgb}{1,0,0}
\newcommand\red[1]{{\color{myred}#1}}

\title{Lie-algebraic classification of effective theories\\ with enhanced soft limits}

\author{Mark P.~Bogers}
\author{and Tom\'a\v{s} Brauner}
\affiliation{Department of Mathematics and Physics, University of Stavanger,\\
N-4036 Stavanger, Norway}
\emailAdd{mark.bogers@uis.no}
\emailAdd{tomas.brauner@uis.no}

\abstract{A great deal of effort has recently been invested in developing methods of calculating scattering amplitudes that bypass the traditional construction based on Lagrangians and Feynman rules. Motivated by this progress, we investigate the long-wavelength behavior of scattering amplitudes of massless scalar particles: Nambu-Goldstone (NG) bosons. The low-energy dynamics of NG bosons is governed by the underlying spontaneously broken symmetry, which likewise allows one to bypass the Lagrangian and connect the scaling of the scattering amplitudes directly to the Lie algebra of the symmetry generators. We focus on theories with \emph{enhanced} soft limits, where the scattering amplitudes scale with a higher power of momentum than expected based on the mere existence of Adler's zero. Our approach is complementary to that developed recently in ref.~\cite{Cheung:2016drk}, and in the first step we reproduce their result. That is, as far as Lorentz-invariant theories with a single physical NG boson are concerned, we find no other nontrivial theories featuring enhanced soft limits beyond the already well-known ones: the Galileon and the Dirac-Born-Infeld (DBI) scalar. Next, we show that in a certain sense, these theories do not admit a nontrivial generalization to non-Abelian internal symmetries. Namely, for compact internal symmetry groups, all NG bosons featuring enhanced soft limits necessarily belong to the center of the group. For noncompact symmetry groups such as the $\gr{ISO}(n)$ group featured by some multi-Galileon theories, these NG bosons then necessarily belong to an Abelian normal subgroup. The Lie-algebraic consistency constraints admit two infinite classes of solutions, generalizing the known multi-Galileon and multi-flavor DBI theories.}


\begin{document}
 
\maketitle


\section{Introduction}
\label{sec:intro}

Recent years have seen a surge of interest in novel computational methods for scattering amplitudes in particle physics~\cite{Kosower:2016cph}. On the practical side, the motivation for these efforts has been provided by current and future particle collider experiments, and the need to bypass the combinatorial explosion that plagues standard perturbation theory based on Lagrangians and Feynman diagrams. More fundamentally, however, the work along this direction has brought to light new structures in quantum field theory, completely invisible to standard perturbative techniques (see refs.~\cite{Elvang:2013cua,Benincasa:2013faa,Cheung:2017pzi} for recent reviews). Different methods to evaluate scattering amplitudes in quantum field theory have thus been put forward, based on recursion relations~\cite{Berends:1987me,Britto:2005fq,Cheung:2015cba,Maniatis:2015kex,Low:2017mlh} as well as other approaches~\cite{Cachazo:2013hca,Baadsgaard:2015twa}.

While the original works focused mostly on gauge theory, more recently the behavior of scattering amplitudes in nonrenormalizable effective field theories (EFTs) for massless scalars --- Nambu-Goldstone (NG) bosons --- has attracted considerable attention, see for instance refs.~\cite{Kampf:2012fn,Cheung:2015ota,Luo:2015tat}. In this case, the asymptotic behavior of scattering amplitudes in the limit of zero energy (the \emph{soft limit}) is of particular interest. Namely, spontaneous symmetry breaking implies, apart from the very existence of NG bosons, that the interactions of NG bosons become weak at low energies. The fact that the scattering amplitude for a process involving a NG boson and an arbitrary number of other particles vanishes in the limit where the NG boson momentum goes to zero (single soft limit), is usually referred to as Adler's zero.\footnote{There are some notable exceptions where Adler's zero is absent though~\cite{Watanabe:2014hca,Huang:2015sla,Rothstein:2017twg}.} Apart from the single soft limit~\cite{Cheung:2014dqa,Low:2014nga,Cachazo:2016njl,Kallosh:2016qvo,Karlsson:2017bfv}, other kinematic regimes such as the double soft limit where the momenta of two participating NG bosons are sent to zero simultaneously~\cite{ArkaniHamed:2008gz,Cachazo:2015ksa,Du:2015esa,Low:2015ogb}, have been investigated.

Let us be concrete and consider a scattering process involving a set of $N$ particles with four-momenta $p_1,\dotsc,p_N$. Now deform the momenta by introducing a scaling parameter $z$ and redefining the momenta $p_i$ to $\tilde p_i(z)$ so that:\\[-3ex]
\begin{itemize}
\itemsep0pt
\item All the particles remain on the mass shell regardless of the value of $z$.
\item Energy and momentum conservation is respected regardless of the value of $z$.
\item The first four-momentum is merely rescaled, $\tilde p_1(z)=zp_1$.
\item The other four-momenta have a nonzero limit, $\lim\limits_{z\to0}\tilde p_i(z)\neq0$ for $i\neq1$.\\[-3ex]
\end{itemize}
The scattering amplitude $\Aa(p_1,\dotsc,p_N)$, once expressed in terms of the modified momenta, can then be symbolically expanded in powers of $z$ as
\begin{equation}
\Aa(\tilde p_1(z),\dotsc,\tilde p_N(z))\propto z^\sigma+\text{terms of higher order in $z$.}
\end{equation}
Provided that the particle with four-momentum $p_1$ is a NG boson, the Adler zero condition requires that $\sigma\geq1$. In the following, we will refer to the soft limit as \emph{enhanced} if $\sigma\geq2$. The question of what values the leading power $\sigma$ can take and how it depends on the given theory has been addressed in a number a recent works~\cite{Cheung:2016drk,Cheung:2014dqa,Bellazzini:2016xrt,Bianchi:2016viy,Padilla:2016mno}. A complete classification of Lorentz-invariant EFTs for a single NG boson from the point of view of scaling of scattering amplitudes was accomplished in ref.~\cite{Cheung:2016drk}. One of our goals in this paper is to provide a complementary viewpoint of the problem, reproducing some results obtained therein, and extending them to theories with multiple NG bosons.

The existence of an enhanced soft limit of scattering amplitudes relies crucially on the presence of symmetry in the system that does not commute with spacetime translations. A prominent example is the Galileon symmetry, see refs.~\cite{Curtright:2012gx,Deffayet:2013lga} for a recent review. In its simplest version with a single scalar $\phi$, this assumes the form $\phi(x)\to\phi(x)+a+b_\mu x^\mu$, where $a$ and $b_\mu$ are constant parameters. However, various generalizations involving more degrees of freedom have been devised~\cite{Deffayet:2010zh,Padilla:2010de,Hinterbichler:2010xn,Trodden:2011xh,Allys:2016hfl}. Another well-known example is the Dirac-Born-Infeld (DBI) scalar, which can be thought of as a fluctuation of a four-dimensional brane embedded into a five-dimensional Minkowski spacetime; see ref.~\cite{deRham:2010eu} for a discussion of a relation between the Galileon and DBI theories. As was shown in ref.~\cite{Cheung:2014dqa}, the enhanced symmetries of the Galileon and DBI theories are responsible for the corresponding enhanced soft limits of scattering amplitudes of the NG mode with $\sigma=2$. There is a special case of the Galileon theory that features a doubly enhanced soft limit with $\sigma=3$; this behavior is now understood to stem from an additional symmetry of the special Galileon action, under which the field $\phi$ shifts by a quadratic function of the coordinate~\cite{Hinterbichler:2015pqa,Noller:2015rea,Novotny:2016jkh}. The possibility of shift symmetries with polynomials of higher orders was investigated in refs.~\cite{Hinterbichler:2014cwa,Griffin:2014bta,Griffin:2015hxa}.

Such enhanced internal symmetry not commuting with spacetime translations cannot be realized by unitary operators on the Hilbert space of the system~\cite{Coleman:1967ad}. In other words, it has to be spontaneously broken, which is obvious in the Galileon and DBI examples. However, it does not give rise to additional NG degrees of freedom; spontaneously broken symmetries with this property are referred to as \emph{redundant}~\cite{Low:2001bw,Watanabe:2013iia}. While redundant symmetries certainly impose a set of nonlinear constraints on the low-energy effective action, it is natural to ask what they imply for the actual observables of the theory, if not the existence of a NG boson. The work of Cheung et al.~\cite{Cheung:2016drk,Cheung:2014dqa} hints at an answer to this question: they imply softening of the scattering amplitudes in the long-wavelength limit. This insight was much needed to push forward our understanding of spontaneous symmetry breaking. While in the case of \emph{uniform} internal symmetries,\footnote{A uniform symmetry is one whose generators commute with spacetime translations~\cite{Watanabe:2011ec}.} we now understand both the classification of NG bosons~\cite{Watanabe:2012hr,Hidaka:2012ym} and the construction of the corresponding EFTs~\cite{Watanabe:2014fva,Brauner:2014ata,Andersen:2014ywa}, the case of nonuniform symmetries has been much less clear.

Our long-term goal is to clarify the general relationship between the presence of redundant symmetries and enhanced soft limits of scattering amplitudes of NG bosons, in both relativistic and nonrelativistic setting. In this paper, we take the first step towards this goal by creating a catalog of theories that admit nontrivial redundant symmetries. Much of the work has already been done by Cheung et al.~\cite{Cheung:2016drk,Cheung:2014dqa}. Namely, they classified constructively all Lorentz-invariant theories for a single massless particle featuring enhanced soft limits, and noticed that a redundant symmetry is present in all cases. They then gave a general argument for the enhanced soft limit based on certain identities for the Noether currents of redundant symmetries~\cite{Brauner:2014aha}. We approach the problem from the opposite end, starting from the symmetry. Our motivation is that physical massless scalars are \emph{always} NG bosons, and that interactions of NG bosons are dictated by the symmetry-breaking pattern. Thus, the effective Lagrangian with all its ambiguities is just a necessary evil for us as well; at the end of the day, there \emph{has to} be a direct connection between the algebra of symmetry generators and the scaling parameter $\sigma$ (see figure~\ref{fig:scheme} for an outline of our basic scheme). Our approach is therefore to classify the extensions of the physical symmetry group by additional redundant generators, admitted by Lie-algebraic constraints. This allows us to set rather stringent constraints on possible extensions of the Galileon and DBI theories to systems with multiple NG bosons.

The plan of our paper is as follows. In section~\ref{sec:method} we explain in some detail the technical steps necessary to generate scattering amplitudes from a given symmetry-breaking pattern. In the following sections, we then work out the classification of theories featuring redundant symmetries, and thus enhanced soft limits.  In section~\ref{sec:abelian}, we first reproduce the results of refs.~\cite{Cheung:2016drk,Cheung:2014dqa} regarding Lorentz-invariant theories with a single massless scalar. This sheds new light on the origin of the ``hidden symmetry'' of the special Galileon. In section~\ref{sec:nonabelian}, we then generalize the construction to theories with multiple NG bosons. We summarize and conclude in section~\ref{sec:conclusions}. For the reader's convenience we collect the list of physically relevant Lie-algebraic structures together with the basic building blocks for invariant actions, found in this paper, in appendix~\ref{app:summary}. Some technical details are relegated to appendices~\ref{app:basis} and~\ref{app:WZterms}.

The basic idea to use Lie-algebraic arguments in order to classify effective theories with enhanced soft limits already appeared in the companion paper~\cite{Bogers:2018kuw}. Therein, we reported briefly our main result concerning the structure of theories with multiple massless scalars, whose scattering amplitudes feature a singly enhanced soft limit. In this paper, we provide most technical details of our work, but also further extend the discussion, allowing for doubly enhanced soft limits.


\section{Methodology}
\label{sec:method}

As was shown in ref.~\cite{Cheung:2016drk}, a redundant symmetry which shifts the NG field by a polynomial of degree $n$ in the coordinate leads generally to an enhanced soft limit with $\sigma=n+1$. Thus, a simply enhanced soft limit ($\sigma=2$) requires a redundant symmetry linear in spacetime coordinates. This in turn amounts to adding a new vector generator, $K^\mu$, of the symmetry algebra that has a nonzero commutator with the generator of spacetime translations, $P^\mu$. A doubly enhanced soft limit ($\sigma=3$) would likewise require adding a rank-two tensor generator, $K^{\mu\nu}$, and so on. Our approach is to simply classify possible extensions of the Lie algebra of generators of the physical symmetry by such additional redundant generators. The precise form of the redundant symmetry transformation is not essential at this stage; indeed it is a \emph{consequence} of our formalism rather than its starting point. This offers certain advantage compared to approaches based on an exhaustive scan of possible transformation rules, polynomial in the fields as well as the coordinates~\cite{Noller:2015rea}.

Once the Lie algebra of symmetry generators is known, we work out the details necessary to generate the actual scattering amplitudes. We do so both to check explicitly the scaling degree $\sigma$ in the low-momentum limit, and to provide a catalog of theories that may later be used by others. In all cases, we work out the symmetry transformation rules and the basic building blocks for the effective Lagrangian using a canonical parameterization for the NG fields. In some cases, we provide explicit expressions for the Lagrangian.

\begin{figure}
\includegraphics[width=\textwidth]{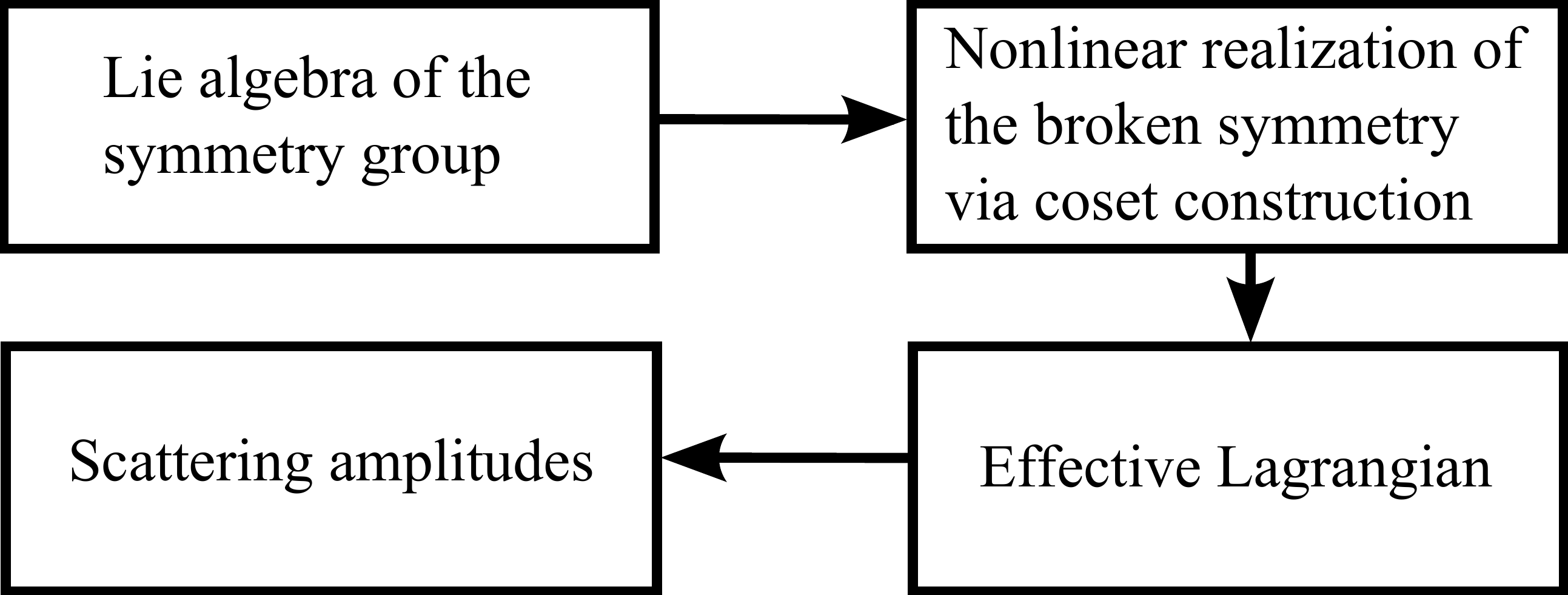}
\caption{The basic scheme of construction of the scattering amplitudes for NG bosons of spontaneously broken symmetry. The scattering amplitudes are fully determined by symmetry except for a few low-energy coupling constants. The latter can in turn be traded for any physical observables, invariant under the reparametrization of the coset space, for instance the values of selected scattering amplitudes at a fixed kinematical point.}
\label{fig:scheme}
\end{figure}

The basic technical steps are outlined in figure~\ref{fig:scheme}. We use the standard method of nonlinear realizations of symmetry, also known as the coset construction~\cite{Callan:1969sn,Volkov:1973vd,Ogievetsky}, which has been widely used to generate effective Lagrangians in both particle physics and cosmology~\cite{Creminelli:2013ygt,Nicolis:2013lma,Creminelli:2014zxa,Goon:2014ika,Delacretaz:2014oxa,Nicolis:2015sra,Zheltukhin:2015jma}. The coset space is generated by all symmetries that are realized nonlinearly, which includes spacetime translations, the broken physical symmetries and the redundant symmetries. One NG field is thereby associated with every broken physical generator and with every redundant generator. Once a parameterization for all the fields has been chosen, the coset construction automatically generates the symmetry transformation rules for us. Then, an invariant action can be constructed solely in terms of a specific set of building blocks, given by the components of the Maurer-Cartan (MC) form and their (covariant) derivatives. For strictly invariant Lagrangians, this is a straightforward procedure using tensor methods, whereas for Lagrangians of the Wess-Zumino (WZ) type, invariant up to a surface term, some extra work is needed~\cite{Witten:1983tw,DHoker:1994ti,DHoker:1995it,Goon:2012dy,Delacretaz:2014jka}. Note that these intermediate steps of the construction of scattering amplitudes necessarily depend on the chosen field parameterization. A possible way out is to focus on reparametrization-invariant quantities that have a well-defined geometrical meaning~\cite{Alonso:2016oah}. In the context of Galileon physics, the freedom to choose the parameterization was behind the discovery of dualities between different Galileon theories~\cite{deRham:2013hsa,Kampf:2014rka}.

\begin{figure}
\includegraphics[width=\textwidth]{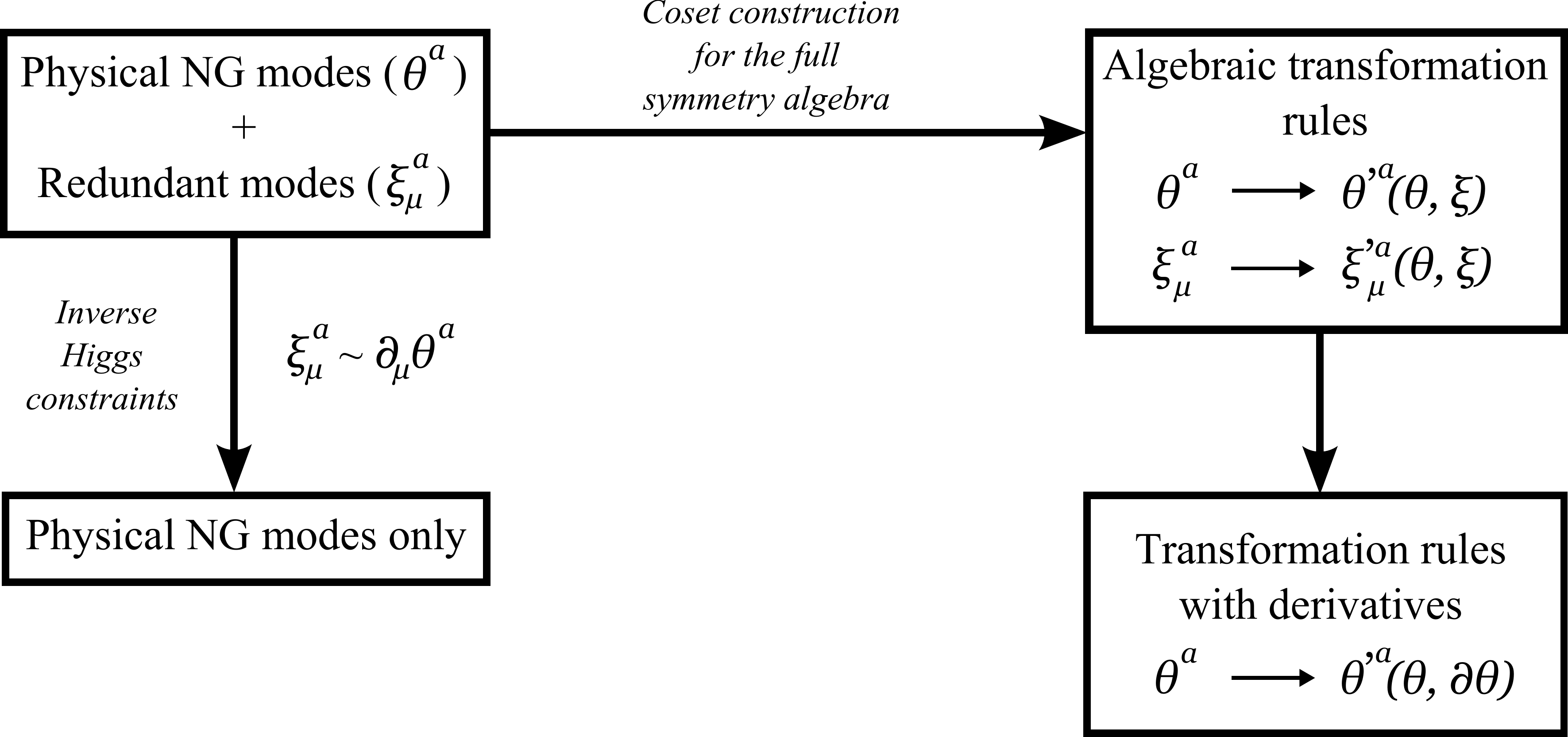}
\caption{The symmetry transformation rules are dictated by symmetry and the choice of parameterization of the coset space, and \emph{never} contain derivatives: they are purely algebraic functions of the coset space coordinates, that is, fields and spacetime coordinates. Transformation rules containing derivatives of the fields can only appear once the redundant modes have been eliminated using a set of inverse Higgs constraints. (The same argument was put forward recently in ref.~\cite{Klein:2017npd}.)}
\label{fig:coset}
\end{figure}

The above-outlined procedure contains a gap though: the redundant symmetries do not give rise to additional physical gapless NG modes in the spectrum. It is now understood that the corresponding modes are either gapped, not being protected by symmetry, or absent from the spectrum altogether~\cite{Endlich:2013vfa,Brauner:2014aha}. Within the EFT based on the coset construction, the fields associated with the redundant generators can be disposed of by an operational prescription known as the \emph{inverse Higgs constraint} (IHC)~\cite{Ivanov:1975zq}. The IHCs are obtained by setting some of the covariant components of the MC form to zero, which ensures their consistency with the symmetry of the system. At this point, it is useful to remark that the symmetry transformation rules generated by the coset construction are always algebraic functions of the fields and spacetime coordinates. The peculiar symmetry of the special Galileon, containing derivatives of the NG field~\cite{Hinterbichler:2015pqa}, is naturally recovered \emph{after} the IHC has been imposed~\cite{Klein:2017npd}, since this dictates the redundant mode to be proportional to the gradient of the physical NG field, see figure~\ref{fig:coset} for a schematic explanation.

\section{Theories with a single NG boson}
\label{sec:abelian}

As the first step, we shall look for relativistic (Poincar\'e-invariant) theories of a single NG boson. The symmetry generators then necessarily include: the generator of spacetime rotations ($J_{\m\n}$), the generator of spacetime translations ($P_\m$), and the generator of the spontaneously broken symmetry ($Q$) that is responsible for the NG boson in the spectrum. In order to fix our conventions, we write down explicitly the already known commutation rules for these generators,
\begin{equation}
\begin{split}
[J_{\m\n},J_{\k\l}]&=\im(g_{\m\l}J_{\n\k}+g_{\n\k}J_{\m\l}-g_{\m\k}J_{\n\l}-g_{\n\l}J_{\m\k}),\\
[J_{\m\n},P_\l]&=\im(g_{\n\l}P_\m-g_{\m\l}P_\n),\\
[J_{\m\n},Q]&=0,\\
[P_\m,P_\n]&=0.
\end{split}
\label{Poincare}
\end{equation}
The first, second and fourth of these just encode the Poincar\'e algebra, whereas the third expresses the fact that $Q$, and thus the NG boson itself, is a Lorentz scalar. We remark that the commutator $[P_\mu,Q]$ is not fixed at this stage: it can be both zero (as for uniform internal symmetries) and nonzero (as, for instance, for spacetime dilatations). To proceed, we need to specify the sector of redundant generators.


\subsection{Simply enhanced soft limit}
\label{subsec:simple}

Should the scattering amplitudes of the NG boson feature an enhanced soft limit with $\sigma=2$, we need, as explained above, an additional vector of redundant generators, $K_\m$. Lorentz invariance requires that the unknown commutators of $K_\m$ with the other generators  as well as the commutator $[P_\m,Q]$ take the following form,\footnote{By inserting terms proportional to $\eps_{\k\l\m\n}$, we restrict ourselves to four spacetime dimensions. However, the coefficients $c$ and $g$ will turn out to be zero, hence the found solutions will apply to Minkowski spacetime of any dimension.}
\begin{align}
\notag
[J_{\m\n},K_\l]&=\im(g_{\n\l}K_\m-g_{\m\l}K_\n),\\
\notag
[P_\m,K_\n]&=\im(\red ag_{\m\n}Q+\red bJ_{\m\n}+\red c\eps_{\m\n\k\l}J^{\k\l}),\\
\label{commU1vector}
[P_\m,Q]&=\im(\red dP_\m+\red eK_\m),\\
\notag
[K_\m,K_\n]&=\im(\red fJ_{\m\n}+\red g\eps_{\m\n\k\l}J^{\k\l}),\\
\notag
[K_\m,Q]&=\im(\red hP_\m+\red iK_\m).
\end{align}
The red-marked unknown coefficients $a$, $b$, $c$, $d$, $e$, $f$, $g$, $h$ and $i$ are constrained by Jacobi identities imposed on the commutators. Once worked out for all possible combinations of generators, these imply that $c=g=0$ and the following additional independent conditions,
\begin{equation}
ae=0,\quad be=0,\quad b+ad=0,\quad b-ai=0,\quad f+ah=0,\quad b(d+i)+ef=0.
\label{JacobiU1vector}
\end{equation}
These conditions have two classes of solutions, depending on whether the coefficient $a$ is zero or nonzero. We will now discuss them in turn; a reader not interested in the details is advised to move on directly to section~\ref{subsec:abeliansummary}, where we summarize the results before we proceed to the construction of the basic building blocks for the effective Lagrangian.


\subsubsection{Unphysical solutions}
\label{subsubsec:unphysical}

The solutions with $a=0$ are ``unphysical'' in that the commutator $[P_\m,K_\n]$ does not contain an admixture of $Q$. This necessarily implies that the NG field for $K^\m$ cannot be eliminated in terms of that for $Q$ by imposing an IHC~\cite{Ivanov:1975zq}. In other words, the generator $K^\m$ is not redundant and does imply the existence of a massless state in the spectrum. This is not the situation we are interested in, we will nevertheless give some details of the solution for the sake of completeness.

The most general solution of the Jacobi identities with $a=0$ reads
\begin{equation}
a=b=c=f=g=0,\qquad\text{$d$, $e$, $h$, $i$ can be arbitrary.}
\end{equation}
The only new nontrivial commutators of the symmetry algebra therefore are
\begin{equation}
[P_\m,Q]=\im(dP_\m+eK_\m),\qquad
[K_\m,Q]=\im(hP_\m+iK_\m).
\end{equation}
These commutators define a linear mapping $X\mapsto[X,Q]$ on the space of generators $X$ with the basis $\{P_\m,K_\m\}$. As such, they can be further simplified by a suitable choice of basis of this space. According to theorem~\ref{thm:22matrix} given in appendix~\ref{app:basis}, one can always find a real basis of the Lie algebra in which the commutation relations take the form
\begin{equation}
[P_\m,Q]=\im(\k P_\m+\l K_\m),\qquad
[K_\m,Q]=\im(s\l P_\m+\k K_\m),
\end{equation}
where $\k$ is real, $\l$ is real non-negative and $s\in\{-1,0,+1\}$. Furthermore, since we have the freedom to rescale the generator $Q$ by an arbitrary nonzero real factor, the final solution for the commutation relations, modulo change of basis, is characterized by a single real parameter and the sign $s$.

While this class of solutions is not relevant for our discussion of soft limits of scattering amplitudes, it may still be of interest to see what geometric structure it corresponds to. We therefore work out the action of various symmetry transformations on spacetime coordinates and fields. It is convenient to parameterize the coset space as
\begin{equation}
U(x,\t,\x)\equiv e^{\im x^\m P_\m}e^{\im\x^\m K_\m}e^{\im\t Q},
\end{equation}
where $\t$ and $\x^\m$ are the NG fields associated with the generators $Q$ and $K^\m$, respectively. Within the coset construction, the transformation properties of all the fields are defined by left multiplication by an element of the symmetry group. This immediately tells us that spacetime translations and transformations generated by $K^\m$ act trivially in that they only shift the coordinate $x^\m$ and the field $\x^\m$, respectively, without affecting the other variables. Finally, to determine the action of the generator $Q$, we have to evaluate the expression $e^{\im\a Q}U$, where $\a$ is the symmetry parameter. A straightforward computation leads to the following result,\footnote{The transformation rule for $\t$ should be read as $\t'(x')=\t(x)+\a$, where $x'^\mu$ is the transformed coordinate. The same remark of course also applies to the transformation of $\x^\m(x)$. The same interpretation of the displayed transformation rules will be assumed implicitly throughout the rest of the paper.}
\begin{equation}
\begin{split}
x^\m&\to e^{\k\a}\bigl[x^\m\cosh(\sqrt s\l\a)+\sqrt s\x^\m\sinh(\sqrt s\l\t)\bigr],\\
\x^\m&\to e^{\k\a}\left[\x^\m\cosh(\sqrt s\l\a)+\frac1{\sqrt s}x^\m\sinh(\sqrt s\l\t)\right],\\
\t&\to\t+\a.
\end{split}
\end{equation}
The transformation rules become particularly simple in the degenerate case $s=0$. Note that for $s=-1$, the hyperbolic functions are simply replaced with the trigonometric ones.


\subsubsection{Physical solutions}
\label{subsec:physicalsol}

Nonzero $a$ implies by means of eq.~\eqref{JacobiU1vector} the following class of solutions,
\begin{equation}
b=ai,\quad c=0,\quad d=-i,\quad e=0,\quad f=-ah,\quad g=0,\quad
\text{$a$, $h$, $i$ can be arbitrary.}
\end{equation}
The nonzero coefficient $a$ can now be eliminated by rescaling $K^\m$ and redefining $h$. Upon renaming the coefficients $h,i$ for the sake of convenience as $u,v$, the nontrivial commutation relations including the internal symmetry generators become
\begin{equation}
\begin{split}
[P_\m,K_\n]&=\im(g_{\m\n}Q+uJ_{\m\n}),\\
[P_\m,Q]&=-\im uP_\m,\\
[K_\m,K_\n]&=-\im vJ_{\m\n},\\
[K_\m,Q]&=\im(vP_\m+uK_\m).
\end{split}
\label{commuv}
\end{equation}
The discussion can be further split into four cases depending on whether the coefficients $u$ and $v$ are zero or nonzero. In all cases, the nonzero coefficient(s) can be eliminated by a rescaling of the generators so that, at the end of the day, the commutation relations contain no free parameters. However, it is convenient to keep the form~\eqref{commuv} since it allows us to switch between the individual cases by taking a suitable deformation of the Lie algebra.

The case of $u=v=0$ corresponds to the Galileon algebra and requires no further discussion. Next, we consider the case $u=0,v\neq0$. Here $v$ can be disposed of by rescaling both $K^\m$ and $Q$ by $\sqrt{|v|}$, leading to the commutation relations
\begin{equation}
[P_\m,K_\n]=\im g_{\m\n}Q,\qquad
[P_\m,Q]=0,\qquad
[K_\m,K_\n]=\mp\im J_{\m\n},\qquad
[K_\m,Q]=\pm\im P_\m,
\label{DBI}
\end{equation}
where the two options correspond to the two signs of $v$. This is equivalent to the five-dimensional Poincar\'e algebra with the four-dimensional momentum and angular momentum operators complemented by $J_{\m4}\equiv K_\m$, $P_4\equiv Q$, and with $g_{44}=\pm 1$. The redundant generators correspond to Lorentz transformations between the four physical spacetime dimension and the fifth dimension, whereas the only physical broken generator is that of the translation in the fifth dimension. The corresponding low-energy EFT describes the fluctuations of a four-brane embedded in a five-dimensional spacetime based on the orthogonal groups $\gr{SO}(3,2)$ and $\gr{SO}(4,1)$, respectively. Both cases correspond to the DBI scalar.

In the $u\neq0,v=0$ case, we can eliminate $u$ by rescaling the $K_\m$ and $Q$ generators, this time regardless of the sign of $u$. The resulting commutation relations read
\begin{equation}
[P_\m,K_\n]=\im(g_{\m\n}Q+J_{\m\n}),\quad
[P_\m,Q]=-\im P_\m,\quad
[K_\m,K_\n]=0,\quad
[K_\m,Q]=\im K_\m.
\label{conformal}
\end{equation}
These are the commutation relations of the conformal group $\gr{SO}(4,2)$ with the dilatation operator $Q$ and the special conformal generator $K_\m$.

Finally, for $u\neq0,v\neq0$, we can eliminate both parameters by rescaling $P^\m$ by $u/\sqrt{|v|}$, $K^\m$ by $\sqrt{|v|}$, and $Q$ by $u$, which results in the set of commutators
\begin{equation}
[P_\m,K_\n]=\im(g_{\m\n}Q+J_{\m\n}),\quad
[P_\m,Q]=-\im P_\m,\quad
[K_\m,K_\n]=\mp\im J_{\m\n},\quad
[K_\m,Q]=\im(K_\m\pm P_\m).
\end{equation}
It is easy to check that upon the redefinition $K^\m\to\tilde K^\m\equiv K^\m\pm\frac{P^\m}2$, this Lie algebra becomes identical to that of eq.~\eqref{conformal}. Hence we are dealing with the conformal algebra $\gr{SO}(4,2)$ again. Regardless of the value of $v$, the $u\neq0$ case therefore corresponds to spontaneous breaking of the conformal group down to the Poincar\'e group. Only the dilatation generator is physical in that it gives rise to a NG mode in the spectrum; the special conformal generators are redundant.

\begin{figure}
\begin{center}
\includegraphics[scale=0.5]{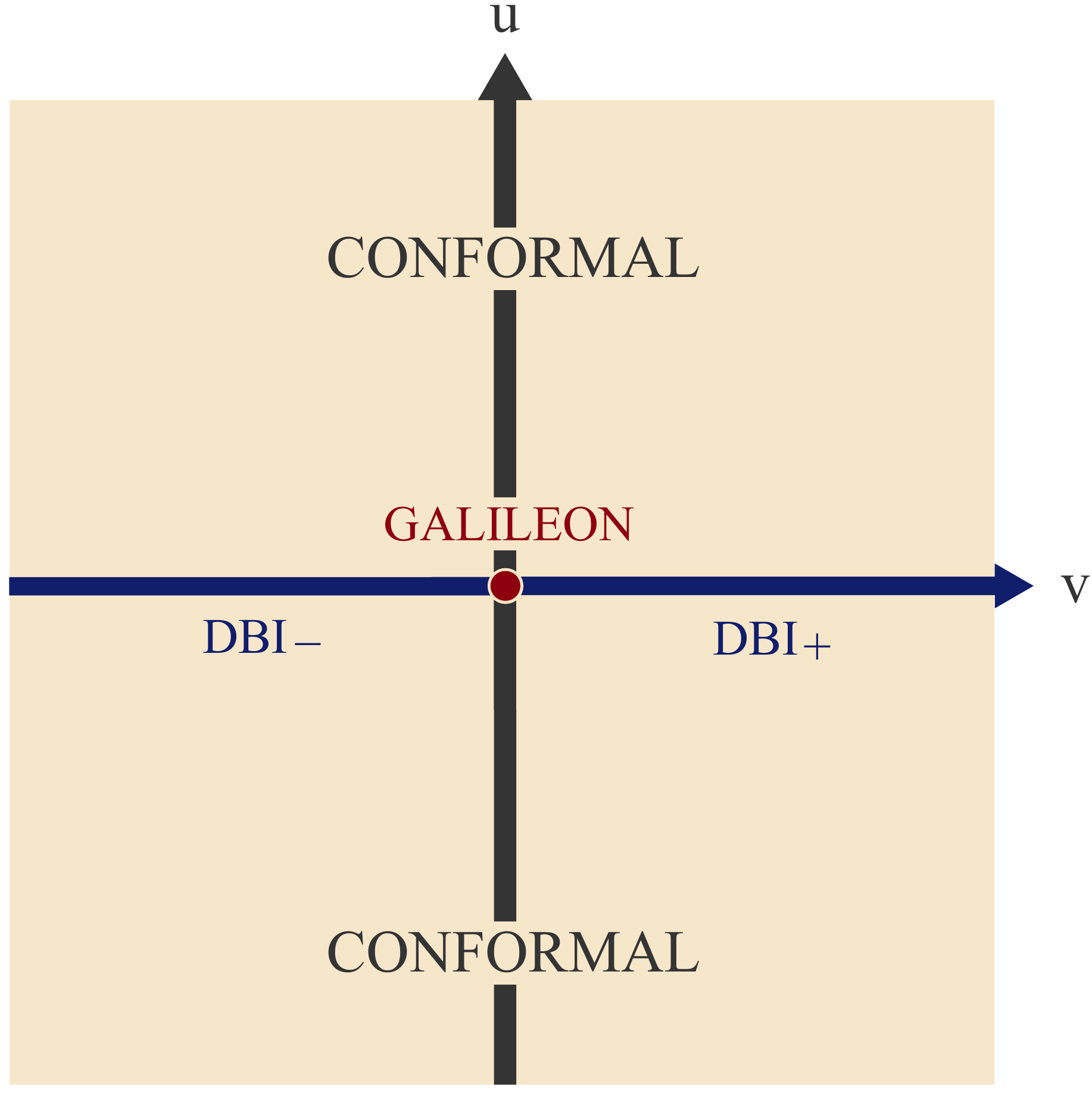}
\end{center}
\caption{The ``phase diagram'' of allowed theories with a single physical NG boson and a vector of redundant generators $K^\m$ as specified by the commutation relations~\eqref{commuv}. The DBI$_+$ and DBI$_-$ symbols refer to DBI theories based on the orthogonal group $\gr{SO}(3,2)$ and $\gr{SO}(4,1)$, respectively.}
\label{fig:conformal}
\end{figure}


\subsubsection{Classification summary}
\label{subsec:abeliansummary}

To summarize, we have classified all possible Poincar\'e-invariant theories with a single physical NG mode and a vector of redundant generators. The nontrivial commutators of the Lie algebra necessarily take the form~\eqref{commuv}. There are three distinct patterns that cannot be transformed to one another by a change of basis of the Lie algebra:\\[-3ex]
\begin{itemize}
\itemsep0pt
\item The Galileon algebra ($u=v=0$).
\item The five-dimensional Poincar\'e algebra ($u=0$, $v\neq0$), leading to the DBI theory.
\item The conformal algebra ($u\neq0$), leading to an EFT for the dilaton.\\[-3ex]
\end{itemize}
The first two are known to lead to enhancement of the soft limit with $\sigma=2$~\cite{Cheung:2014dqa}. The last one, however, does not. This is consistent with the fact that the dilatation generator does not commute with $P^\m$, which may spoil the soft limit~\cite{Watanabe:2014hca,Rothstein:2017twg}. In the following, we will therefore take $[P_\m,Q]=0$ as an additional assumption when trying to map out possible theories with enhanced soft limits.

Note that the three different cases are related by simple deformations, tuning the values of the parameters $u,v$, as is clear from eq.~\eqref{commuv}. This is represented graphically by the ``phase diagram'' in  figure~\ref{fig:conformal}. We now use the formulation of eq.~\eqref{commuv} to work out the basic building blocks for effective Lagrangians that can be used for all three cases.


\subsubsection{Coset construction of effective Lagrangians}
\label{subsec:coset}

Following the straightforward algorithm of the coset construction, we first have to choose a parameterization for the coset element. A convenient choice in this case is
\begin{equation}
U(x,\t,\x)\equiv e^{\im x^\m P_\m}e^{\im\t Q}e^{\im\x^\m K_\m},
\end{equation}
where $\t$ is the physical NG field, whereas $\x^\m$ is the ``would-be'' NG mode, excited by the redundant generator $K^\m$. The MC one-form is defined as usual by
\begin{equation}
\om\equiv-\im U^{-1}\dd U.
\end{equation}
It can be decomposed into components corresponding to the individual generators of the symmetry group,
\begin{equation}
\om=\frac12\om_J^{\m\n}J_{\m\n}+\om_P^\m P_\m+\om_K^\m K_\m+\om_Q Q.
\end{equation}
A straightforward, if slightly tedious, calculation gives
\begin{align}
\notag
\om_J^{\m\n}={}&-ue^{u\t}\frac{\sin\sqrt{v\x^2}}{\sqrt{v\x^2}}(\x^\n\dd x^\m-\x^\m\dd x^\n)+\frac1{\x^2}(1-\cos\sqrt{v\x^2})(\x^\n\dd\x^\m-\x^\m\dd\x^\n),\\
\notag
\om_P^\m={}&e^{u\t}\dd x^\m+e^{u\t}\frac{\x^\m\x\cdot\dd x}{\x^2}(\cos\sqrt{v\x^2}-1)+v\x^\m\dd\t\frac{\sin\sqrt{v\x^2}}{\sqrt{v\x^2}},\\
\label{wideeq}
\om_K^\m={}&\frac uve^{u\t}\left(2\frac{\x^\m\x\cdot\dd x}{\x^2}-\dd x^\m\right)(\cos\sqrt{v\x^2}-1)+u\x^\m\dd\t\frac{\sin\sqrt{v\x^2}}{\sqrt{v\x^2}}\\
\notag
&+\dd\x^\m+\left(\frac{\x^\m\x\cdot\dd\x}{\x^2}-\dd\x^\m\right)\biggl(1-\frac{\sin\sqrt{v\x^2}}{\sqrt{v\x^2}}\biggr),\\
\notag
\om_Q={}&-e^{u\t}\x\cdot\dd x\frac{\sin\sqrt{v\x^2}}{\sqrt{v\x^2}}+\dd\t\cos\sqrt{v\x^2}.
\end{align}
The unphysical redundant mode $\x^\m$ can be eliminated by imposing the IHC $\om_Q=0$, which is equivalent to
\begin{equation}
\x_\m\frac{\tan\sqrt{v\x^2}}{\sqrt{v\x^2}}=e^{-u\t}\de_\m\t.
\label{IHC}
\end{equation}
The physical meaning of the other components of the MC form is then as follows. The $\om^\m_P$ represents a covariant vielbein, defined through $\om^\a_PP_\a\equiv e^\a_\m\dd x^\m P_\a$, and is, among others, needed to construct an invariant volume measure for spacetime integrals. The $\om^{\m\n}_J$ is a spin connection, used to define covariant derivatives of fields with nonzero spin. Finally, $\om^\m_K$ contains the covariant derivative of the $\x^\m$ field, which is the basic building block for construction of invariant Lagrangians.

Upon using the IHC~\eqref{IHC}, the covariant vielbein acquires the form\footnote{Note that we sometimes use the IHC~\eqref{IHC} to express $\t$ in terms of $\x^\m$ rather than vice versa: it leads to more compact expressions. At the end of the day, the unphysical field $\x^\m$ is to be solved for using eq.~\eqref{IHC}.}
\begin{equation}
e^\a_\m=e^{u\t}\biggl(\Pe_{\perp\m}^\a+\frac1{\cos\sqrt{v\x^2}}\Pe^\a_{\parallel\m}\biggr),
\end{equation}
where $\Pe_\perp$ and $\Pe_\parallel$ are projectors to directions perpendicular and parallel to $\x^\m$, defined as
\begin{equation}
\Pe^\m_{\perp\n}\equiv\delta^\m_\n-\frac{\x^\m\x_\n}{\x^2},\qquad
\Pe^\m_{\parallel\n}\equiv\frac{\x^\m\x_\n}{\x^2}.
\end{equation}
The induced metric on the coset space then becomes\footnote{We use the capitalized symbol $G_{\m\n}$ to distinguish the metric on the coset space, pulled back to the Minkowski spacetime, from the physical, flat-space Minkowski metric $g_{\m\n}$.}
\begin{equation}
G_{\m\n}\equiv g_{\a\b}e^\a_\m a^\b_\n=e^{2u\t}g_{\m\n}+v\de_\m\t\de_\n\t.
\label{metric}
\end{equation}
This makes the interpretation of $\t$ as the dilaton in the case $u\neq0$ obvious. The covariant derivative of the $\x^\m$ field, defined through $\om^\a_K\equiv e^\a_\n\dd x^\m\nabla_\m\x^\n$, is obtained likewise as
\begin{equation}
\nabla_\m\x^\n=\frac{u}{v}(1-\cos\sqrt{v\x^2})\delta_\m^\n+\biggl(\Pe^\n_{\perp\a}\frac{\sin\sqrt{v\x^2}}{\sqrt{v\x^2}}+\Pe^\n_{\parallel\a}\cos\sqrt{v\x^2}\biggr)e^{-u\t}\de_\m\x^\a.
\end{equation}
Invariant actions can now be obtained using the volume measure $\dd^4x\sqrt{-G}$, multiplied by a Lagrangian constructed solely out of $\nabla_\m\x^\n$ and its covariant derivatives. The zeroth-order action thus takes the form
\begin{equation}
S_0=\int\dd^4x\sqrt{-G}=\int\dd^4x\,e^{4u\t}\sqrt{1+ve^{-2u\t}\de_\m\t\de^\m\t},
\end{equation}
and the corresponding flat-space Lagrangian in the three cases of interest reads
\begin{equation}
\begin{alignedat}{3}
\La_0&=1&&\text{Galileon }(u=v=0),\\
&=\sqrt{1+v\de_\m\t\de^\m\t}\qquad\qquad&&\text{DBI }(u=0,v\neq0),\\
&=e^{4u\t}&&\text{conformal }(u\neq0,v=0).
\end{alignedat}
\end{equation}
For illustration, we also display the next contribution to the action, which reads
\begin{equation}
\begin{split}
S_1={}&\int\dd^4x\sqrt{-G}\,\nabla_\m\x^\m\\
={}&\int\dd^4x\,e^{4u\t}\biggl\{\frac{4u}v\Bigl(\sqrt{1+ve^{-2u\t}\de_\m\t\de^\m\t}-1\Bigr)\\
&+\sqrt ve^{-2u\t}\biggl[\de_\m\de^\m\t-u(\de_\m\t)^2-ve^{-2u\t}\frac{\de^\m\t\de^\n\t(\de_\m\de_\n\t-u\de_\m\t\de_\n\t)}{1+ve^{-2u\t}\de_\m\t\de^\m\t}\biggr]\biggr\}.
 \end{split}
\end{equation}
The corresponding flat-space Lagrangians then become
\begin{equation}
\begin{alignedat}{3}
\La_1&=0&&\text{Galileon }(u=v=0),\\
&=\sqrt v\biggl(\de_\m\de^\m\t-v\frac{\de^\m\t\de^\n\t\de_\m\de_\n\t}{1+v\de_\m\t\de^\m\t}\biggr)\qquad\qquad&&\text{DBI }(u=0,v\neq0),\\
&=2ue^{2u\t}\de_\m\t\de^\m\t&&\text{conformal }(u\neq0,v=0).
\end{alignedat}
\end{equation}
The above-found Lagrangians for the DBI and conformal cases agree with those found in ref.~\cite{Goon:2012dy}, whereas the ``Galileon'' Lagrangians are trivial: the actual action of the Galileon theories is determined using the WZ construction upon setting $u=v=0$~\cite{Goon:2012dy}. It is, however, also possible to obtain the Galileon Lagrangians by setting $u=0$, expanding the action in powers of $v$, and then picking the coefficient in front of $v^1$~\cite{deRham:2010eu,Bogers:2018kuw}.

At this point, it is useful to remark that in line with the general philosophy of EFT~\cite{Weinberg:1996v2}, we consider a given theory to be fully determined by the corresponding symmetry structure. Its Lagrangian therefore as a rule contains an infinite tower of operators, whose couplings are to be determined by experiment or by matching to an underlying microscopic theory. This convention is somewhat different from the literature where, for instance, the term ``Galileon'' refers to a Lagrangian with merely a \emph{finite} number of operators, namely only those with a number of derivatives per field smaller than $\sigma=2$. In our terminology, even an operator such as, say, $(\Box\t)^4(\de_\m\de_\n\t)^2$, belongs to the Galileon action, being allowed by its symmetry. It is, of course, only the operators with a sufficiently low number of derivatives that provide a nontrivial realization of the enhanced soft limits. Our Lie-algebraic approach suggests a straightforward way to classify such \emph{exceptional} EFTs in the sense of ref.~\cite{Cheung:2016drk}. The same remark applies to all the other EFTs constructed in the rest of this paper.

To conclude our discussion, we finally work out the transformation properties of all the coset fields under the symmetry group, which sheds more light on the nature of the three systems of interest here. First, spacetime translations only affect the coordinate $x^\m$ and merely shift it in the expected manner. To work out the field transformations under the internal symmetry generated by $Q$, we have to evaluate $e^{\im\a Q}U$, where $\a$ is the parameter of the transformation. It is straightforward to show that $\x^\m$ remains intact whereas
\begin{equation}
x^\m\to x^\m e^{-u\a},\qquad
\t\to\t+\a.
\end{equation}
The rescaling of the coordinate agrees with the fact that for $u\neq0$, the symmetry algebra coincides with the conformal algebra and $Q$ plays the role of the dilatation operator.

The transformation generated by the operator $K^\m$ is likewise obtained by multiplying $U$ by the matrix $e^{\im\b^\m K_\m}$ from the left. After some algebra, we find
\begin{equation}
\begin{split}
x^\m&\to x^\m+u\b\cdot xx^\m-\frac12ux^2\b^\m-\frac vue^{-u\t}\b^\m\sinh u\t+\mathcal O(\b^2),\\
\t&\to\t+\b\cdot x+\mathcal O(\b^2),\\
\x^\m&\to\x^\m+e^{-u\t}\b^\m+u\x\cdot x\b^\m-u\x\cdot\b x^\m+\mathcal O(\beta^2,\x^2).
\end{split}
\label{transfoAbelian}
\end{equation}
Note that while the first two rules are exact expressions valid to first order in a power expansion in $\b^\m$, the last line requires a Taylor expansion in the redundant field $\x^\m$ as well. More complete expressions can be obtained in the special case $u=0$ which is of most interest as in this case, the scattering amplitudes actually feature soft limits. Here we find after a straightforward albeit somewhat tedious computation that
\begin{equation}
\begin{split}
x^\m&\to x^\m-\frac{\sin\sqrt{v\b^2}}{\sqrt{v\b^2}}v\t\b^\m+(\cos\sqrt{v\b^2}-1)\frac{\b^\m\b_\n}{\b^2}x^\n,\\
\t&\to\t\cos\sqrt{v\b^2}+\b\cdot x\frac{\sin\sqrt{v\b^2}}{\sqrt{v\b^2}},\\
\x^\m&\to\x^\m+\biggl(\Pe^\m_{\perp\n}\frac{\sqrt{v\x^2}}{\tan\sqrt{v\x^2}}+\Pe^\m_{\parallel\n}\biggr)\b^\n+\mathcal O(\b^2).
\end{split}
\end{equation}


\subsection{Doubly enhanced soft limit}
\label{subsec:double}

Now that we have mapped out all Lorentz-invariant theories with a single physical NG boson that could feature enhanced soft limits, we address the question which of them might possess even softer amplitudes with $\sigma=3$. In accord with the general argument of ref.~\cite{Cheung:2016drk}, this requires adding another set of redundant operators, generating symmetry transformations quadratic in the coordinates. This means adding a rank-two tensor generator, which allows for different options, corresponding in three spatial dimensions to spin zero, one and two, respectively.


\subsubsection{Spin-zero multiplet of redundant generators}
\label{subsec:spinzero}

We start our discussion with the simplest case of spin zero. In other words, we assume that the algebra of the generators $J^{\m\n}$, $P^\m$, $Q$ and $K^\m$, discussed in section~\ref{subsec:simple}, is complemented by another scalar generator, denoted as $X$. Lorentz invariance fixes, in addition to eq.~\eqref{Poincare}, the following commutators,
\begin{equation}
\begin{split}
[J_{\m\n},K_\l]&=\im(g_{\n\l}K_\m-g_{\m\l}K_\n),\\
[J_{\m\n},X]&=0.
\end{split}
\end{equation}
The remaining commutators of the Lie algebra can be parameterized by a set of numerical coefficients, similarly to eq.~\eqref{commU1vector}. Keeping the notation for the $a,\dotsc,i$ terms, introduced therein, we write down the most general Lie-algebraic structure admitted by Lorentz invariance as
\begin{align}
\notag
[P_\m,K_\n]&=\im(\red ag_{\m\n}Q+\red bJ_{\m\n}+\red c\eps_{\m\n\k\l}J^{\k\l}+\red jg_{\m\n}X),\\
\notag
[P_\m,Q]&=\im(\red dP_\m+\red eK_\m),\\
\notag
[K_\m,K_\n]&=\im(\red fJ_{\m\n}+\red g\eps_{\m\n\k\l}J^{\k\l}),\\
\label{commU1vectorX}
[K_\m,Q]&=\im(\red hP_\m+\red iK_\m),\\
\notag
[P_\m,X]&=\im(\red kP_\m+\red\ell K_\m),\\
\notag
[K_\m,X]&=\im(\red mP_\m+\red nK_\m),\\
\notag
[Q,X]&=\im(\red oQ+\red pX).
\end{align}
The full set of constraints following from the Jacobi identities would be too long to even write down here. We therefore focus on the special case of $a\neq0$ and $\ell\neq0$; these conditions are required to make the generators $K^\m$ and $X$ redundant, and thus to be able to eliminate the corresponding ``would-be'' NG fields. With these assumptions, one can readily solve for the unknown parameters $a,\dotsc,p$, getting
\begin{equation}
\begin{gathered}
b=0,\qquad c=0,\qquad d=\frac{ek}\ell,\qquad f=0,\qquad g=0,\qquad h=\frac{em}\ell,\qquad i=\frac{en}\ell,\\
j=-\frac{ae}\ell,\qquad o=k+n,\qquad p=-\frac e\ell(k+n),
\end{gathered}
\end{equation}
the remaining parameters $a,e,k,\ell,m,n$ being free. The resulting commutation relations look somewhat involved, but they simplify dramatically upon the subsequent redefinition $Q\to\tilde Q\equiv Q-\frac e\ell X$. This brings the Lie algebra to the form (dropping the tilde)
\begin{equation}
\begin{split}
[P_\m,K_\n]&=\im ag_{\m\n}Q,\\
[P_\m,X]&=\im(kP_\m+\ell K_\m),\\
[K_\m,X]&=\im(mP_\m+nK_\m),\\
[Q,X]&=\im(k+n)Q,
\end{split}
\end{equation}
with $[P_\m,Q]=[K_\m,Q]=[K_\m,K_\n]=0$. Thus, out of all the Lie algebraic structures with a singly enhanced soft limit, presented in section~\ref{subsec:physicalsol}, only the Galileon algebra admits extension by adding the scalar $X$.

The commutations relations can be further simplified by utilizing theorem~\ref{thm:22matrix} displayed in appendix~\ref{app:basis}. Upon absorbing the nonzero parameter $a$ in the redefinition of $Q$, and the parameter $\l$, introduced in appendix~\ref{app:basis} and required to be nonzero for $X$ to be redundant, into the redefinition of $X$, the nontrivial commutators of the Lie algebra take the final form
\begin{equation}
\begin{split}
[P_\m,K_\n]&=\im g_{\m\n}Q,\\
[P_\m,X]&=\im(\k P_\m+K_\m),\\
[K_\m,X]&=\im(sP_\m+\k K_\m),\\
[Q,X]&=2\im\k Q,
\end{split}
\end{equation}
where $\k$ is a real parameter that can be both zero and nonzero, and $s\in\{-1,0,+1\}$. We expect that in analogy with the conformal algebra case, the parameter $\kappa$ should vanish for the scattering amplitudes to feature Adler's zero.

We start the analysis of the found Lie algebra by working out the transformations rules for the fields. The coset element will be defined as
\begin{equation}
U(x,\t,\x,\phi)\equiv e^{\im x^\m P_\m}e^{\im\t Q}e^{\im\x^\m K_\m}e^{\im\phi X}.
\end{equation}
The spacetime translation acts trivially in that it merely shifts the coordinate $x^\m$ and does not affect the NG fields $\t,\x^\m,\phi$. Likewise, the transformation generated by $Q$ acts trivially in that it only shifts the $\t$ field. The transformation generated by $K^\m$, $e^{\im\b^\m K_\m}$, acts just like the linear shift of the Galileon symmetry,
\begin{equation}
\t\to\t+\b\cdot x,\qquad
\x^\m\to\x^\m+\b^\m.
\end{equation}
To work out the transformation generated by $X$, we multiply $U$ by $e^{\im\om X}$. A straightforward calculation then leads to
\begin{equation}
\begin{split}
x^\m&\to e^{\k\om}\bigl[x^\m\cosh(\sqrt s\om)+\sqrt s\x^\m\sinh(\sqrt s\om)\bigr],\\
\t&\to e^{2\k\om}\left[\t+\frac12\left(\frac{x^2}{\sqrt s}+\x^2\sqrt s\right)\sinh(\sqrt s\om)\cosh(\sqrt s\om)+\x\cdot x\sinh^2(\sqrt s\om)\right],\\
\x^\m&\to e^{\k\om}\left[\x^\m\cosh(\sqrt s\om)+\frac1{\sqrt s}x^\m\sinh(\sqrt s\om)\right],\\
\phi&\to\phi+\om.
\end{split}
\end{equation}
Of particular interest is the case $s=0$, making the transformation rules extremely simple,
\begin{equation}
x^\m\to e^{\k\om}x^\m,\qquad
\t\to e^{2\k\om}\left(\t+\frac12\om x^2\right),\qquad
\x^\m\to e^{\k\om}(\x^\m+\om x^\m),\qquad
\phi\to\phi+\om.
\end{equation}
This case corresponds to a trivial extension of the linear Galileon shift of the NG field $\t$ by a shift quadratic in the coordinate, possibly twisted by an overall dilatation.

The effective action can be constructed using the MC one-form, $\om\equiv-\im U^{-1}\dd U$, which can in this case be decomposed as
\begin{equation}
\om=\om_P^\m P_\m+\om_K^\m K_\m+\om_QQ+\om_XX.
\end{equation}
A straightforward manipulation gives the result,
\begin{equation}
\begin{split}
\om^\m_P&=e^{-\k\phi}\bigl[\dd x^\m\cosh(\sqrt s\phi)-\sqrt s\,\dd\x^\m\sinh(\sqrt s\phi)\bigr],\\
\om^\m_K&=e^{-\k\phi}\left[\dd\x^\m\cosh(\sqrt s\phi)-\frac1{\sqrt s}\,\dd x^\m\sinh(\sqrt s\phi)\right],\\
\om_Q&=e^{-2\k\phi}(\dd\t-\x\cdot\dd x),\\
\om_X&=\dd\phi.
\end{split}
\label{MCspin0}
\end{equation}
Let us see if this MC form can be used to construct theories where enhanced soft limits of scattering amplitudes are realized nontrivially. The auxiliary field $\x^\m$ can be eliminated by setting $\om_Q=0$, which leads to $\x_\m=\de_\m\t$. Likewise, the auxiliary field $\phi$ has to be eliminated by choosing a suitable IHC. Since $\om^\m_P$ serves as a vielbein, the only choice compatible with Lorentz invariance is to project out the singlet component of the covariant derivative of $\x^\m$, stemming from $\om^\m_K$, that is, $\nabla_\m\x^\m$. This will make $\phi$ a function of the gradient of $\x^\m$. In short, upon imposing the IHCs, $\x^\m$ carries one derivative acting on $\t$, while $\phi$ carries two derivatives acting on $\t$.

The effective action is now to be constructed out of the remaining components of the MC form: the vielbein $e^\a_\m$ stemming from $\om^\m_P$, the traceless part of $\nabla_\m\x^\n$ and $\om_X=\de_\m\phi\,\dd x^\m$. The $e^\a_\m$ and $\nabla_\m\x^\n$ only depend on the gradient of $\x^\m$ and on $\phi$, and hence carry two derivatives per every factor of $\t$. $\om_X$ is a derivative of $\phi$, which itself is a function of the gradient of $\x^\m$. To conclude, every invariant Lagrangian constructed out of the MC form carries at least two derivatives per every factor of $\t$. The symmetry constraints therefore do not allow us to even construct a standard kinetic term $\t$. This means that there are no theories of an interacting massless scalar which would realize the assumed Poincar\'e symmetry, extended by the redundant vector $K^\m$ and Lorentz singlet $X$.

In the above argument, we have only taken into account the strictly invariant terms in the effective Lagrangian, constructed directly out of local products of the MC form and its covariant derivatives. Another possibility to get an invariant effective action is through the WZ construction which, after all, is how the Galileon action is obtained in the singly enhanced case~\cite{Goon:2012dy}. This is done in appendix~\ref{app:WZterms}. We find, however, no nontrivial terms that can contribute to the action for a physical massless scalar.


\subsubsection{Spin-one multiplet of redundant generators}

Next, we want to see if it is possible to extend the algebra of generators $J^{\m\n}$, $P^\m$, $Q$ and $K^\m$ by an additional antisymmetric tensor $A^{\m\n}$. Lorentz invariance implies that, in addition to eq.~\eqref{Poincare}, the following commutators are fixed,
\begin{equation}
\begin{split}
[J_{\m\n},K_\l]&=\im(g_{\n\l}K_\m-g_{\m\l}K_\n),\\
[J_{\m\n},A_{\k\l}]&=\im(g_{\m\l}A_{\n\k}+g_{\n\k}A_{\m\l}-g_{\m\k}A_{\n\l}-g_{\n\l}A_{\m\k}).
\end{split}
\end{equation}
The remaining commutators of the symmetry Lie algebra can be parameterized by a priori unknown coefficients $a,\dotsc,y$ as
\begin{align}
\notag
[P_\m,K_\n]={}&\im(\red ag_{\m\n}Q+\red bJ_{\m\n}+\red c\eps_{\m\n\k\l}J^{\k\l}+\red jA_{\m\n}+\red r\eps_{\m\n\k\l}A^{\k\l}),\\
\notag
[P_\m,Q]={}&\im(\red dP_\m+\red eK_\m),\\
\notag
[K_\m,K_\n]={}&\im(\red fJ_{\m\n}+\red g\eps_{\m\n\k\l}J^{\k\l}+\red sA_{\m\n}+\red t\eps_{\m\n\k\l}A^{\k\l}),\\
\notag
[K_\m,Q]={}&\im(\red hP_\m+\red iK_\m),\\
\notag
[A_{\m\n},A_{\k\l}]={}&\im\bigl[\red k(g_{\m\l}J_{\n\k}+g_{\n\k}J_{\m\l}-g_{\m\k}J_{\n\l}-g_{\n\l}J_{\m\k})\\
\label{commU1antisym}
&+\red\ell(g_{\m\l}\eps_{\n\k\a\b}+g_{\n\k}\eps_{\m\l\a\b}-g_{\m\k}\eps_{\n\l\a\b}-g_{\n\l}\eps_{\m\k\a\b})J^{\a\b}\\
\notag
&+\red u(g_{\m\l}A_{\n\k}+g_{\n\k}A_{\m\l}-g_{\m\k}A_{\n\l}-g_{\n\l}A_{\m\k})\\
\notag
&+\red v(g_{\m\l}\eps_{\n\k\a\b}+g_{\n\k}\eps_{\m\l\a\b}-g_{\m\k}\eps_{\n\l\a\b}-g_{\n\l}\eps_{\m\k\a\b})A^{\a\b}\bigr],\\
\notag
[A_{\m\n},P_\l]={}&\im\bigl[\red m(g_{\m\l}P_\n-g_{\n\l}P_\m)+\red n(g_{\m\l}K_\n-g_{\n\l}K_\m)\bigr],\\
\notag
[A_{\m\n},K_\l]={}&\im\bigl[\red o(g_{\m\l}P_\n-g_{\n\l}P_\m)+\red p(g_{\m\l}K_\n-g_{\n\l}K_\m)\bigr],\\
\notag
[A_{\m\n},Q]={}&\im(\red qA_{\m\n}+\red w\eps_{\m\n\k\l}A^{\k\l}+\red xJ_{\m\n}+\red y\eps_{\m\n\k\l}J^{\k\l}).
\end{align}
We assume right away that $a\neq0$ in order that the vector $K^\m$ is redundant. An explicit solution of the constraints following from the Jacobi identities then shows that $n=o=0$ and $m=p$. This implies that the new tensor $A^{\m\n}$ cannot be redundant, independently of the choice of basis in the subspace of generators spanned on $P^\m$ and $K^\m$. We conclude that this scenario is not viable: there is no extension of the Poincar\'e algebra augmented with the scalar $Q$ by an additional antisymmetric tensor $A^{\m\n}$, in which it would be redundant.


\subsubsection{Spin-two multiplet of redundant generators}

Finally, we focus on the spin-two case. We will thus assume that apart from the Poincar\'e group generators, the symmetry algebra of the system contains the physical broken generator $Q$, a redundant vector $K^\m$, and a redundant traceless symmetric tensor $S^{\m\n}$. Lorentz invariance dictates, in addition to eq.~\eqref{Poincare}, the following commutation relations,
\begin{equation}
\begin{split}
[J_{\m\n},K_\l]&=\im(g_{\n\l}K_\m-g_{\m\l}K_\n),\\
[J_{\m\n},S_{\k\l}]&=\im(-g_{\m\l}S_{\n\k}+g_{\n\k}S_{\m\l}-g_{\m\k}S_{\n\l}+g_{\n\l}S_{\m\k}).
\end{split}
\end{equation}
The remaining commutation relations of the symmetry algebra can be parameterized by a set of numerical coefficients, similarly to eq.~\eqref{commU1vector}. Lorentz invariance and the tracelessness of $S^{\m\n}$ dictates the following structure in four spacetime dimensions,
\begin{equation}
\begin{split}
[P_\m,K_\n]={}&\im(\red ag_{\m\n}Q+\red bJ_{\m\n}+\red c\eps_{\m\n\k\l}J^{\k\l}+\red jS_{\m\n}),\\
[P_\m,Q]={}&\im(\red dP_\m+\red eK_\m),\\
[K_\m,K_\n]={}&\im(\red fJ_{\m\n}+\red g\eps_{\m\n\k\l}J^{\k\l}),\\
[K_\m,Q]={}&\im(\red hP_\m+\red iK_\m),\\
[S_{\m\n},S_{\k\l}]={}&\im[\red k(g_{\m\l}J_{\n\k}+g_{\n\k}J_{\m\l}+g_{\m\k}J_{\n\l}+g_{\n\l}J_{\m\k})\\
&+\red\ell(g_{\m\l}\eps_{\n\k\a\b}+g_{\n\k}\eps_{\m\l\a\b}+g_{\m\k}\eps_{\n\l\a\b}+g_{\n\l}\eps_{\m\k\a\b})J^{\a\b}],\\
[S_{\m\n},P_\l]={}&\im\bigl[\red m(g_{\m\l}P_\n+g_{\n\l}P_\m-\tfrac12g_{\m\n}P_\l)+\red n(g_{\m\l}K_\n+g_{\n\l}K_\m-\tfrac12g_{\m\n} K_\l)\bigr],\\
[S_{\m\n},K_\l]={}&\im\bigl[\red o(g_{\m\l}P_\n+g_{\n\l}P_\m-\tfrac12g_{\m\n}P_\l)+\red p(g_{\m\l}K_\n+g_{\n\l}K_\m-\tfrac12g_{\m\n} K_\l)\bigr],\\
[S_{\m\n},Q]={}&\im\red qS_{\m\n},
\end{split}
\label{commU1tensor}
\end{equation}
where we used the same notation for the $a,\dotsc,i$ terms as in eq.~\eqref{commU1vector} and labeled as $j,\dotsc,q$ the new terms in the commutators. Again, the full set of constraints following from the Jacobi identities would be too long to write down explicitly. We therefore restrict ourselves from the outset to the special case of $a\neq0$. With this assumption, the solution for the parameters of the Lie algebra can be given as
\begin{equation}
\begin{gathered}
b=jm,\quad c=0,\quad d=-\frac{5jm}{2a},\quad e=0,\quad f=jo,\quad g=0,\quad h=-\frac{5jo}{2a},\\
i=\frac{5jm}{2a},\quad k=m^2+no,\quad\ell=0,\quad p=-m,\quad q=0,
\end{gathered}
\end{equation}
where $a$ is arbitrary nonzero, $m$ and $o$ are arbitrary, and $j$ and $n$ must satisfy the constraint $jn=0$. The commutation relations~\eqref{commU1tensor} thereby reduce to
\begin{align}
\notag
[P_\m,K_\n]={}&\im(ag_{\m\n}Q+jmJ_{\m\n}+jS_{\m\n}),\\
\notag
[P_\m,Q]={}&-\im\frac{5jm}{2a}P_\m,\\
\notag
[K_\m,K_\n]={}&\im joJ_{\m\n},\\
[K_\m,Q]={}&\im\frac{5j}{2a}(-oP_\m+mK_\m),\\
\notag
[S_{\m\n},S_{\k\l}]={}&\im(m^2+no)(g_{\m\l}J_{\n\k}+g_{\n\k}J_{\m\l}+g_{\m\k}J_{\n\l}+g_{\n\l}J_{\m\k}),\\
\notag
[S_{\m\n},P_\l]={}&\im\bigl[m(g_{\m\l}P_\n+g_{\n\l}P_\m-\tfrac12g_{\m\n}P_\l)+n(g_{\m\l}K_\n+g_{\n\l}K_\m-\tfrac12g_{\m\n} K_\l)\bigr],\\
\notag
[S_{\m\n},K_\l]={}&\im\bigl[o(g_{\m\l}P_\n+g_{\n\l}P_\m-\tfrac12g_{\m\n}P_\l)-m(g_{\m\l}K_\n+g_{\n\l}K_\m-\tfrac12g_{\m\n} K_\l)\bigr],\\
\notag
[S_{\m\n},Q]={}&0.
\end{align}
We will now argue that physically interesting solutions can only exist if $j=0$. Let us assume otherwise for the sake of contradiction, which implies $n=0$. The very existence of Adler's zero, which translates to the condition $[P_\m,Q]=0$, then in addition requires $m=0$. But then the generator $S^{\m\n}$ is not redundant since $[S_{\m\n},P_\l]=0$, and the corresponding massless tensor NG mode remains in the spectrum, which is not the situation we are interested in.\footnote{Note that this conclusion is independent of the choice of basis in the $\{P^\m,K^\m\}$ subspace of generators. Namely, a redefinition $\{P^\m,K^\m\}\to\{\tilde P^\m,\tilde K^\m\}$ that would allow the commutator $[S_{\m\n},\tilde P_\l]$ to contain an admixture of $\tilde K_\m$ requires nonzero $o$, yet such a redefinition would also lead to nonzero $[\tilde P_\m,\tilde P_\n]$, which would be in contradiction with the interpretation of $\tilde P^\m$ as the generator of spacetime translations.} Once we know that $j=0$, the only nontrivial commutators that contain the redundant generators are
\begin{equation}
\begin{split}
[P_\m,K_\n]={}&\im ag_{\m\n}Q,\\
[S_{\m\n},S_{\k\l}]={}&\im(m^2+no)(g_{\m\l}J_{\n\k}+g_{\n\k}J_{\m\l}+g_{\m\k}J_{\n\l}+g_{\n\l}J_{\m\k}),\\
[S_{\m\n},P_\l]={}&\im\bigl[m(g_{\m\l}P_\n+g_{\n\l}P_\m-\tfrac12g_{\m\n}P_\l)+n(g_{\m\l}K_\n+g_{\n\l}K_\m-\tfrac12g_{\m\n} K_\l)\bigr],\\
[S_{\m\n},K_\l]={}&\im\bigl[o(g_{\m\l}P_\n+g_{\n\l}P_\m-\tfrac12g_{\m\n}P_\l)-m(g_{\m\l}K_\n+g_{\n\l}K_\m-\tfrac12g_{\m\n} K_\l)\bigr],
\end{split}
\end{equation}
where $a,m,n,o$ can take arbitrary values except that $a$ and $n$ are nonzero. We conclude that out of all the Lie-algebraic structures with a singly enhanced soft limit, discussed in section~\ref{subsec:physicalsol}, only the Galileon algebra admits extension by adding the tensor $S^{\m\n}$.

Now that $[K_\m,K_\n]=0$, we can further simplify the coefficients of the Lie algebra by changing basis in the space spanned on $\{P^\m,K^\m\}$. Using theorem~\ref{thm:22matrix} from appendix~\ref{app:basis}, the commutation relations can by a suitable choice of basis be simplified to
\begin{equation}
\begin{split}
[P_\m,K_\n]={}&\im g_{\m\n}Q,\\
[S_{\m\n},S_{\k\l}]={}&\im s(g_{\m\l}J_{\n\k}+g_{\n\k}J_{\m\l}+g_{\m\k}J_{\n\l}+g_{\n\l}J_{\m\k}),\\
[S_{\m\n},P_\l]={}&\im(g_{\m\l}K_\n+g_{\n\l}K_\m-\tfrac12g_{\m\n} K_\l),\\
[S_{\m\n},K_\l]={}&\im s(g_{\m\l}P_\n+g_{\n\l}P_\m-\tfrac12g_{\m\n}P_\l),
\end{split}
\label{commtraceless}
\end{equation}
where $s\in\{-1,0,+1\}$. (The parameter $\lambda$ of theorem~\ref{thm:22matrix} can be eliminated by rescaling $S^{\m\n}$.)

Next, we introduce a parameterization for the coset element,
\begin{equation}
U(x,\t,\x,\b)\equiv e^{\im x^\m P_\m}e^{\im\t Q}e^{\im\x^\m K_\m}e^{\frac\im2\b^{\m\n}S_{\m\n}},
\end{equation}
where $\b^{\m\n}$ is a traceless symmetric tensor of auxiliary fields, corresponding to the generator $S^{\m\n}$. The transformation properties of all the fields are as usual defined by left multiplication of $U$ by an element of the symmetry group. Spacetime translations and the transformations generated by $Q$ and $K^\m$ act in the same way as in the spin-zero case, worked out in section~\ref{subsec:spinzero}. The transformation generated by $S^{\m\n}$ is obtained by multiplying $U$ by $e^{\frac\im2\om^{\m\n}S_{\m\n}}$. A straightforward calculation then leads to
\begin{equation}
\begin{split}
x^\m&\to x_\n\cosh(\sqrt s\om)^{\m\n}-\sqrt s\x_\n\sinh(\sqrt s\om)^{\m\n},\\
\t&\to\t-\frac12\left(\frac{x_\m x_\n}{\sqrt s}+\x_\m\x_\n\sqrt s\right)\bigl[\sinh(\sqrt s\om)\cosh(\sqrt s\om)\bigr]^{\m\n}+\x_\m x_\n\bigl[\sinh^2(\sqrt s\om)\bigr]^{\m\n},\\
\x^\m&\to\x_\n\cosh(\sqrt s\om)^{\m\n}-\frac1{\sqrt s}x_\n\sinh(\sqrt s\om)^{\m\n},\\
\b^{\m\n}&\to\b^{\m\n}+\om^{\m\n}+\mathcal O(\om^2,\b^2).
\end{split}
\label{transfotraceless}
\end{equation}

Invariant actions are constructed with the MC form, whose components are defined by
\begin{equation}
\om=\frac12\om_J^{\m\n}J_{\m\n}+\om_P^\m P_\m+\frac12\om_S^{\m\n}S_{\m\n}+\om_K^\m K_\m+\om_Q Q.
\end{equation}
With the shorthand notation
\begin{equation}
B^{\a\b}_{\m\n}\equiv\b^\a_\m\delta^\b_\n-\b^\b_\n\delta^\a_\m,
\end{equation}
these components take the form
\begin{equation}
\begin{split}
\om^{\m\n}_J&=\dd\b^{\a\b}\bigl\{B^{-1}[\cosh(\sqrt sB)-\mathbbm1]\bigr\}^{\m\n}_{\a\b},\\
\om^\m_P&=\dd x_\n\cosh(\sqrt s\b)^{\m\n}+\sqrt s\,\dd\x_\n\sinh(\sqrt s\b)^{\m\n},\\
\om^{\m\n}_S&=\dd\b^{\a\b}\frac{\bigl[B^{-1}\sinh(\sqrt sB)\bigr]^{\m\n}_{\a\b}}{\sqrt s},\\
\om^\m_K&=\dd x_\n\frac{\sinh(\sqrt s\b)^{\m\n}}{\sqrt s}+\dd\x_\n\cosh(\sqrt s\b)^{\m\n},\\
\om_Q&=\dd\t-\x\cdot\dd x.
\end{split}
\label{tracelessMCform}
\end{equation}
The auxiliary field $\x^\m$ is eliminated by setting $\om_Q=0$, which corresponds to $\x_\m=\de_\m\t$. Note that eq.~\eqref{transfotraceless} then implies the following transformation rule for the physical field $\t$,
\begin{equation}
\label{transfortracelessIHC}
\t\to\t-\frac12\om^{\m\n}(x_\m x_\n+s\de_\m\t\de_\n\t)+\mathcal O(\om^2).
\end{equation}
This naturally recovers the ``hidden symmetry'' of the special Galileon, first reported in ref.~\cite{Hinterbichler:2015pqa}. The corresponding Lie algebra found therein matches our eq.~\eqref{commtraceless}. The special case of $s=0$ then corresponds to a mere shift of the NG field $\t$, quadratic in the spacetime coordinates~\cite{Hinterbichler:2014cwa}. 

The construction of WZ terms in this spin-two case is again reviewed in appendix~\ref{app:WZterms}. It turns out that there is a single WZ term that can contribute to the action of a physical massless scalar regardless of the value of $s$. For $s=\pm1$, this reproduces the special Galileon, whereas for $s=0$, it is just the kinetic term, $(\de_\m\t)^2$; it is easy to see that this changes by a surface term upon a traceless quadratic shift of the field, defined by eq.~\eqref{transfortracelessIHC}.

What other, strictly invariant operators can be constructed out of the MC form? In order to eliminate the auxiliary field $\b^{\m\n}$, we need a symmetric traceless rank-two tensor of IHCs, which is naturally chosen as the symmetric traceless part of $\nabla_\m\x_\n$. What is left of the MC form after all IHCs have been applied is the singlet covariant derivative $\nabla_\m\x^\m$, the antisymmetric part of $\nabla_\m\x_\n$, and the covariant derivative of $\b^{\m\n}$, represented by $\om^{\m\n}_S$. By the same argument as in the spin-zero case, these contain two or more derivatives per every factor of $\t$. This says that for $s=\pm1$, the special Galileon gives a leading contribution to the action. However, invariant Lagrangians constructed from the above-listed building blocks may still give interactions that realize nontrivially doubly enhanced soft limits of scattering amplitudes.

The $s=0$ case is somewhat different in that the WZ term is a pure noninteracting kinetic term, and thus all interactions, if present, must come from the invariant part of the Lagrangian. Setting $s=0$, eq.~\eqref{tracelessMCform} becomes
\begin{equation}
\om^{\m\n}_J=0,\quad
\om^\m_P=\dd x^\m,\quad
\om^{\m\n}_S=\dd\b^{\m\n},\quad
\om^\m_K=\dd x_\n\b^{\m\n}+\dd\x^\m,\quad
\om_Q=\dd\t-\x\cdot\dd x.
\end{equation}
Imposing the IHCs leads to
\begin{equation}
\x_\m=\de_\m\t,\qquad
\b_{\m\n}=-\frac12\left(\de_\m\x_\n+\de_\m\x_\m-\frac12g_{\m\n}\de_\a\x^\a\right)=-\de_\m\de_\n\t+\frac14g_{\m\n}\Box\t.
\end{equation}
The remaining nonzero components of the MC form then correspond to
\begin{equation}
\nabla_\m\x^\m=\Box\t,\qquad
\nabla_\l\b_{\m\n}=-\de_\m\de_\n\de_\l\t+\frac14g_{\m\n}\de_\l\Box\t;
\end{equation}
the antisymmetric part of $\nabla_\m\x_\n=\de_\m\de_\n\t+\b_{\m\n}$ vanishes identically. These ingredients do not lead to any interesting theory though. The operator $\nabla_\l\b_{\m\n}$ contains three derivatives and thus gives a trivial doubly enhanced soft limit when acting on an asymptotic state of a scattering process. The operator $\nabla_\m\x^\m$, on the other hand, cannot even act on any asymptotic massless state, as it would give a strict zero on the mass shell. In fact, theories with Lagrangian of the type
\begin{equation}
\La=\frac12(\de_\m\t)^2+f(\Box\t),
\end{equation}
where $f$ is an arbitrary analytic function of its variable, are easily seen to be equivalent to the theory of a free massless scalar upon a suitable field redefinition.


\section{Theories with multiple NG bosons}
\label{sec:nonabelian}

In the previous section, we analyzed in detail theories with a single NG boson type (\emph{flavor}), just to confirm the results of previous works that there are no nontrivial systems featuring enhanced soft limits beyond the Galileon and DBI theories. However, we used these examples to work out a systematic method to classify candidate theories, which we would now like to generalize to cases with more than one physical NG boson. It is of course trivially possible to, for instance, simply add two copies of the Galileon to get a theory with two NG bosons and enhanced soft limits. What we would like to see, however, is whether, and to what extent, enhanced soft limits can be found in systems where a \emph{non-Abelian} internal symmetry is spontaneously broken.

We therefore introduce a set of internal symmetry generators $Q_i$, and correspondingly a set of redundant vector generators $K^\m_A$. The most general set of commutation relations for these generators and the Poincar\'e generators $J^{\m\n}$ and $P^\m$, allowed by Lorentz invariance, reads in analogy with eqs.~\eqref{Poincare} and~\eqref{commU1vector},
\begin{equation}
\begin{split}
[J_{\m\n},J_{\k\l}]&=\im(g_{\m\l}J_{\n\k}+g_{\n\k}J_{\m\l}-g_{\m\k}J_{\n\l}-g_{\n\l}J_{\m\k}),\\
[J_{\m\n},P_\l]&=\im(g_{\n\l}P_\m-g_{\m\l}P_\n),\\
[J_{\m\n},K_{\l A}]&=\im(g_{\n\l}K_{\m A}-g_{\m\l}K_{\n A}),\\
[J_{\m\n},Q_i]&=0,\\
[P_\m,P_\n]&=0,\\
[P_\m,K_{\n A}]&=\im(\red{a^i_A}g_{\m\n}Q_i+\red{b_A}J_{\m\n}+\red{c_A}\eps_{\m\n\k\l}J^{\k\l}),\\
[P_\m,Q_i]&=\im(\red{d_i}P_\m+\red{e^A_i}K_{\m A}),\\
[K_{\m A},K_{\n B}]&=\im(\red{f_{AB}}J_{\m\n}+\red{g_{AB}}\eps_{\m\n\k\l}J^{\k\l}+\red{\hh^i_{AB}}g_{\m\n}Q_i),\\
[K_{\m A},Q_i]&=\im(\red{h_{Ai}}P_\m+\red{i^B_{Ai}}K_{\m B}),\\
[Q_i,Q_j]&=\im\red{\str^k_{ij}}Q_k.
\end{split}
\label{commnonAvector}
\end{equation}
In order to facilitate a comparison with the case of a single NG flavor, we again kept the notation for the $a,\dotsc,i$ terms in the commutators, and denoted as $\hh,\str$ the new, genuinely multi-flavor contributions.\footnote{For the same reason, we are unfortunately unable to maintain the same notation as in ref.~\cite{Bogers:2018kuw}. The final results as reviewed in appendix~\ref{app:summary} are, however, free from this ambiguity.} The coefficients $f_{AB}$ and $g_{AB}$ are symmetric in their indices, while $\hh^i_{AB}$ is antisymmetric under $A\leftrightarrow B$ and $\str^k_{ij}$ is antisymmetric under $i\leftrightarrow j$.

It is straightforward to work out the constraints on the Lie algebra imposed by the Jacobi identities. However, we will not attempt to find a general solution. Instead, we will focus on the case where the $Q_i$s generate a uniform symmetry and we can thus expect a soft limit featuring Adler's zero~\cite{Watanabe:2014hca,Rothstein:2017twg}. In other words, we set
\begin{equation}
d_i\to0,\qquad
e^A_i\to0.
\end{equation}
The set of Jacobi identities then reduces to
\begin{equation}
b_A=0,\qquad c_A=0,\qquad f_{AB}=-a^i_Ah_{Bi},\qquad g_{AB}=0,
\end{equation}
accompanied by the independent constraints
\begin{align}
\label{eqai}
a^i_A i^C_{Bi}&=0,\\
\label{eqail}
a^k_Bi^B_{Ai}+a^j_A\str^k_{ij}&=0,\\
\label{eqah}
a^k_Ah_{Bi}-a^k_Bh_{Ai}+i^C_{Ai}\hh^k_{BC}-i^C_{Bi}\hh^k_{AC}&=\str^k_{ij}\hh^j_{AB},\\
\label{eqih}
i^B_{Ai}h_{Bj}-i^B_{Aj}h_{Bi}&=\str^k_{ij}h_{Ak},\\
\label{eqii}
i^C_{Ai}i^B_{Cj}-i^C_{Aj}i^B_{Ci}&=\str^k_{ij}i^B_{Ak},\\
\label{eqll}
\str^m_{ij}\str^n_{mk}+\str^m_{jk}\str^n_{mi}+\str^m_{ki}\str^n_{mj}&=0,\\
\label{eqhxi}
h_{Ai}\hh^i_{BC}&=0,\\
\label{eqfxi}
f_{AC}\delta^D_B-f_{BC}\delta^D_A&=\hh^i_{AB}i^D_{Ci}.
\end{align}
Some of these have an obvious group-theoretic interpretation. For instance, eq.~\eqref{eqll} is the usual Jacobi identity for the set of internal generators $Q_i$. Likewise, eq.~\eqref{eqii} says that the matrices $(\Ka_i)^A_{\phantom AB}\equiv-\im i^A_{Bi}$ furnish a representation of the Lie algebra of $Q_i$. Moreover, according to eq.~\eqref{eqai}, the linear combinations $a^i_A\Ka_i$ vanish in this representation. In fact, there is a simple geometric interpretation of all the above constraints, which allows one to construct symmetry algebras with redundant generators in terms of certain vector space endowed with an invariant metric, and an affine representation of the internal symmetry generators $Q_i$ on this space. This geometric picture is elaborated in Ref.~\cite{Bogers:2018kuw}; here we focus on working out \emph{explicitly} a class of theories ready-made for applications in cosmology and high-energy physics.


\subsection{Single redundant generator}

To warm up, we first consider the case where there is a single redundant vector $K^\m$ but multiple broken generators. Physically, we expect this to correspond to a theory with several NG bosons, out of which only one has enhanced soft limits of its scattering amplitudes. We can now abandon the index $A$ and the Jacobi constraints simplify drastically,
\begin{equation}
a^ii_i=0,\qquad a^ki_i+a^j\str^k_{ij}=0,\qquad i_ih_j-i_jh_i=\str^k_{ij}h_k,\qquad\str^k_{ij}i_k=0,
\label{oneK}
\end{equation}
along with the Jacobi identity for $\str^k_{ij}$~\eqref{eqll} and the explicit solution $f=-a^ih_i$. Note that the first condition is not independent as long as \emph{some} $a^i$ is nonzero, which we anyway assume since otherwise $K^\m$ would not be a redundant generator: by contracting the second condition with $a^i$, we get $a^ka^ii_i=0$ for any $k$, from which the first follows. 

Let us introduce the compact notation $\tilde Q\equiv a^iQ_i$ and $v\equiv a^ih_i$. Then the nontrivial commutators in eq.~\eqref{commnonAvector}, including the redundant generators, can be split into two classes,
\begin{equation}
\begin{aligned}{}
[P_\m,K_\n]&=\im g_{\m\n}\tilde Q,\\
[K_\m,K_\n]&=-\im vJ_{\m\n},\\
[K_\m,\tilde Q]&=\im vP_\m,
\end{aligned}
\qquad\qquad
\begin{aligned}{}
[K_\m,Q_i]&=\im(h_iP_\m+i_iK_\m),\\
[Q_i,Q_j]&=\im\str^k_{ij}Q_k.
\end{aligned}
\label{generalized_DBI}
\end{equation}
The commutators in the first class are identical to the special case $u=0$ of eq.~\eqref{commuv}, whereas the commutators in the second class take into account the possibly non-Abelian nature of the internal symmetry. We can therefore think of the present Lie algebra as a generalization of either the DBI or the Galileon system to several internal symmetry generators, and it is natural to split our following discussion accordingly in the two scenarios.


\subsubsection{DBI-like systems}
\label{subsec:DBI}

Let us therefore first assume that $v\neq0$. It then follows at once that $i_i=0$ by multiplying the second condition in eq.~\eqref{oneK} by $h_k$ and using the other conditions therein.\footnote{The same conclusion can be reached even without assuming $v\neq0$ for compact semisimple Lie algebras. Namely, in such cases there is a positive-definite invariant metric that can be used to raise and lower adjoint indices. Moreover, the rank-three tensor $\str_{ijk}$ is fully antisymmetric~\cite{Andersen:2014ywa}. Contracting the second condition in eq.~\eqref{oneK} with $a_k$ and using this antisymmetry then leads to $a_ka^ki_i=0$, which by the positivity of the metric implies that $i_i=0$ as long as some $a^i$ is nonzero.} The set of constraints~\eqref{oneK} then boils down to
\begin{equation}
a^j\str^k_{ij}=0,\qquad h_k\str^k_{ij}=0.
\label{ahconstr}
\end{equation}
Let us take $\tilde Q$ as one of the generators and redefine the other generators as $\tilde Q_i\equiv Q_i-\frac{h_i}v\tilde Q$. Then, the above constraints on $a^i$ and $h_i$ together with all the commutation relations can be encoded in the following set of conditions,
\begin{equation}
\begin{aligned}{}
[P_\m,K_\n]&=\im g_{\m\n}\tilde Q,\qquad\qquad
&[K_\m,\tilde Q_i]&=0,\\
[K_\m,K_\n]&=-\im vJ_{\m\n},
&[\tilde Q,\tilde Q_i]&=0,\\
[K_\m,\tilde Q]&=\im vP_\m,
&[\tilde Q_i,\tilde Q_j]&=\im\str^k_{ij}\tilde Q_k.
\end{aligned}
\end{equation}
The resulting generalization of the DBI theory is to a large extent trivial: this symmetry algebra is a direct sum of the DBI algebra, discussed in section~\ref{subsec:physicalsol}, and the algebra of the non-Abelian generators $\tilde Q_i$. The construction of basic building blocks of invariant actions for the DBI part can be copy-pasted from section~\ref{subsec:coset}. We will therefore only briefly review the coset construction for the internal algebra of the $\tilde Q_i$s, following the classic paper~\cite{Callan:1969sn}. The coset element is parameterized as
\begin{equation}
\tilde U(\tilde\t)\equiv e^{\im\tilde\t^a\tilde Q_a}.
\label{DBI_param}
\end{equation}
We use the index notation of ref.~\cite{Andersen:2014ywa} in which $\tilde Q_{i,j,\dotsc}$ stands for a generic generator of the internal symmetry group, $\tilde Q_{\a,\b,\dotsc}$ for an unbroken one, and $\tilde Q_{a,b,\dotsc}$ for a broken one. The MC form for such broken internal symmetry will be denoted as
\begin{equation}
\Omega\equiv-\im e^{-\im\tilde\t\cdot\tilde Q}\dd e^{\im\tilde\t\cdot\tilde Q}.
\label{Omegaform}
\end{equation}
The transformation rules for the NG fields $\tilde\t^a$ are defined by left multiplication of the coset element $\tilde U(\tilde\t)$ by a group element $g$,
\begin{equation}
g\tilde U(\tilde\t)=\tilde U(\tilde\t')h(g,\tilde\t),
\label{gtransfo}
\end{equation}
where $h$ belongs to the unbroken subgroup and in general depends on both $g$ and the NG fields $\tilde\t^a$. We can parameterize it exponentially as $h=e^{\im k^\a\tilde Q_\a}$ . While there is no closed expression for $k^\a(g,\tilde\t)$, it can be calculated order by order in the NG fields $\tilde\t^a$, at least for $g$ infinitesimally close to unity. Under the transformation~\eqref{gtransfo}, the MC form~\eqref{Omegaform} then changes as
\begin{equation}
\Omega\to h\Omega h^{-1}-\im h\dd h^{-1}.
\label{MCtransfo}
\end{equation}
Since $h$ belongs to the unbroken subgroup, the unbroken component of $\Omega$, $\Omega^\a\tilde Q_\a$ transforms as a gauge field of the unbroken subgroup and can be used to construct covariant derivatives of operators transforming in any linear representation of this subgroup. The broken component $\Omega^a\tilde Q_a$, on the other hand, transforms covariantly under the adjoint action of the unbroken subgroup. It defines the covariant derivatives $\nabla_\m\tilde\t^a$ of the NG fields $\tilde\t^a$, and constitutes the basic building block of invariant actions.


\subsubsection{Galileon-like systems}

The generalization of the Galileon theory, based on the Lie algebra~\eqref{generalized_DBI}, is characterized by $v=0$. In this case, we cannot provide an explicit general solution to the conditions~\eqref{oneK} as in the generalized DBI case. Yet, the commutation relations~\eqref{generalized_DBI} simplify dramatically. We can then evaluate the MC form explicitly with the single additional technical assumption that the generators $Q_i$ can, just like in the DBI-like case, be split into $\tilde Q\equiv a^iQ_i$ and $\tilde Q_i$ such that the latter form a closed Lie algebra, $[\tilde Q_i,\tilde Q_j]=\im\str^k_{ij}\tilde Q_k$. Denoting the NG fields associated with $\tilde Q$ and $\tilde Q_a$ as $\t$ and $\tilde\t^a$, respectively, it is then convenient to use the following parameterization of the coset space,
\begin{equation}
U(x,\t,\tilde\t,\x)\equiv e^{\im x^\m P_\m}e^{\im\x^\m K_\m}e^{\im\t\tilde Q}e^{\im\tilde\t^a\tilde Q_a}.
\end{equation}
A straightforward calculation then leads to an expression for the MC form
\begin{equation}
\begin{split}
\om_P^\m&=\dd x^\m+h_a\tilde\t^a\frac{e^{-i_b\tilde\t^b}-1}{i_c\tilde\t^c}\dd\x^\m,\\
\om^\m_K&=e^{-i_a\tilde\t^a}\dd\x^\m,\\
\om_{\tilde Q}&=e^{-i_a\tilde\t^a}(\dd\t-\x\cdot\dd x).
\end{split}
\label{GalileonMC}
\end{equation}
The MC form for the internal generators $\tilde Q_i$ is given by $\Omega$, defined in eq.~\eqref{Omegaform}. The exponential factors in the other components of the MC form arise from the commutator $[\tilde Q_i,\tilde Q]=-\im i_i\tilde Q$, which follows from the properties of the Lie algebra coefficients~\eqref{oneK}.

As explained below eq.~\eqref{MCtransfo}, the broken components of the MC form transform covariantly under the adjoint action of the unbroken subgroup. The presence of $P_\m$ in the commutator $[K_\m,\tilde Q_i]$ then implies that only if $h_\a=0$, $\omega_P^\m$ is invariant under the internal symmetry, and hence can serve as a covariant vielbein.\footnote{There was no such restriction in the previously discussed case of DBI-like systems where $[K_\m,\tilde Q_i]=0$.} We expect to eliminate the redundant mode $\x^\m$ by imposing the IHC $\om_{\tilde Q}=0$. Just like in the simplest Galileon theory, discussed in section~\ref{sec:abelian}, all the remaining components of the MC form then depend on the second derivative of $\t$. In order to generate a kinetic term for $\t$, and thus have a well-defined perturbative dynamics, we have to resort to the WZ construction. However, it seems that there are no nontrivial Lie-algebraic structures that would admit WZ terms constructed from the MC form~\eqref{GalileonMC}, see appendix~\ref{app:WZterms} for more details.

We therefore have to broaden our search for interesting theories with nontrivially realized enhanced soft limits of scattering amplitudes. We do so by extending our scope to systems with multiple redundant generators.


\subsection{Multiple redundant generators}

A general solution to all the Jacobi constraints on the symmetry Lie algebra, subject only to the assumption that $[P_\m,Q_i]=0$, is given in Ref.~\cite{Bogers:2018kuw}. Here we will work out in detail two particular, infinite classes of solutions, generalizing the Galileon and DBI systems, discussed in section~\ref{subsec:physicalsol}.


\subsubsection{Generalized Galileon solutions}

We start with the class of Lie algebras for which
\begin{equation}
h_{Ai}=0,\qquad
\hh^i_{AB}=0.
\end{equation}
The former assumption necessarily implies that $f_{AB}=0$, and hence ensures that the generators $Q_i$ and $K_{\m A}$ form a closed Lie algebra and thus generate a truly internal symmetry. The latter, technical assumption ensures that $Q_i$ and $K_{\m A}$ separately form closed Lie algebras. Introducing the set of generators $Q_A\equiv a^i_AQ_i$, all the remaining nontrivial constraints among eqs.~\eqref{eqai}--\eqref{eqfxi} are then encoded in the commutation relations
\begin{equation}
\begin{split}
[P_\m,K_{\n A}]&=\im g_{\m\n}Q_A,\\
[Q_i,K_{\m A}]&=(\Ka_i)^B_{\phantom BA}K_{\m B},\\
[Q_i,Q_A]&=(\Ka_i)^B_{\phantom BA}Q_B,\\
[Q_A,Q_B]&=0,
\end{split}
\label{general_Galileon}
\end{equation}
together with $[K_{\m A},K_{\n B}]=0$ and the condition that the matrices $(\Ka_i)^B_{\phantom BA}$ define a representation of the internal symmetry subgroup generated by $Q_i$.

The generators $Q_A$ appearing in $[P_\m,K_{\n A}]$ obviously define an Abelian ideal of the full internal symmetry algebra generated by all the $Q_i$s. For compact internal symmetry groups, the $Q_A$s then necessarily belong to the center of the symmetry algebra, and the corresponding NG bosons featuring enhanced soft limits are thus associated to one or more $\gr{U}(1)$ factors of the symmetry group. Once the compactness requirement is relaxed, other possibilities exist. An example is provided by one of the multi-flavor generalizations of the Galileon theory, where the $Q_i$s span the Lie algebra $\gr{ISO}(n)$~\cite{Goon:2012dy}. The physical NG modes correspond to the $n$ mutually commuting translations, transforming as a vector under the $\gr{SO}(n)$ rotations, which remain unbroken. In this case, the internal symmetry group $\gr{ISO}(n)$ is non-semisimple, being isomorphic to the semidirect product $\gr{SO}(n)\ltimes\mathbb{R}^n$.

Note that the general class of Lie algebras, describing multiple NG bosons with enhanced soft limits, defined by eq.~\eqref{general_Galileon}, is pretty robust. All one has to do to specify such a theory uniquely is to choose the Lie algebra for the generators $Q_i$ and its Abelian ideal, generated by $Q_A$. There are no more arbitrary parameters involved in the construction; all other commutation relations are then fixed by eq.~\eqref{general_Galileon}.

Further simplification arises in the special case that, similarly to the $\gr{ISO}(n)$-symmetric multi-Galileon theory, the generators $Q_i$ can be split into subsets, $\tilde Q_i$ and $Q_A$, such that the $\tilde Q_i$s themselves generate a closed Lie algebra. The Lie algebra of $Q_i$ is then a semidirect sum of the two subalgebras, generated by $\tilde Q_i$ and $Q_A$.\footnote{Examples of Lie-algebraic structures of the type~\eqref{general_Galileon} that do \emph{not} satisfy this assumption can easily be constructed though, the simplest one being the Heisenberg algebra. More generally, one can consider central extensions of the algebra of $Q_i$, where the central charges naturally belong among the Abelian generators $Q_A$. We are indebted to Torsten Schoeneberg and Qiaochu Yuan for clarifying this point to us.} It can be constructed algorithmically as follows. Take any Lie algebra $\mathfrak{g}$ with the generators $\tilde Q_i$ and its real, finite-dimensional representation $R$; let $n$ be the dimension of this representation. Treating $\mathbb{R}^n$ as an Abelian Lie algebra and $Q_A$ as its generators, construct the full Lie algebra of $Q_i$ as the semidirect sum $\mathfrak{g}\ltimes\mathbb{R}^n$, where the action of $\mathfrak{g}$ on $\mathbb{R}^n$ is defined by the representation $R$.

All the multi-Galileon theories, constructed so far in the literature, are of this latter type, where the generators $\tilde Q_i$ define the group $\gr{SO}(n)$ or $\gr{SU}(n)$ and the generators $Q_A$ its fundamental or adjoint representation~\cite{Hinterbichler:2010xn,Trodden:2011xh,Padilla:2010ir}. In these examples, all the generators $\tilde Q_i$ remain unbroken so that the only NG modes in the system are the Galileon ones, associated with the spontaneously broken shift symmetries, generated by $Q_A$. However, that in general does not have to be the case. As we will now see, the construction of systems based on a semidirect sum algebra $\mathfrak{g}\ltimes\mathbb{R}^n$ can be carried out in full detail for an arbitrary Lie algebra $\mathfrak{g}$ and its arbitrary real finite-dimensional representation $R$, regardless of which generators of $\mathfrak{g}$ are spontaneously broken and which are not.

To work out the basic building blocks for the effective Lagrangians, we first introduce the following parameterization of the coset space in analogy with eq.~\eqref{DBI_param},
\begin{equation}
U(x,\t,\x)\equiv e^{\im x^\m P_\m}e^{\im\t^AQ_A}e^{\im\x^{\m A}K_{\m A}}e^{\im\t^a\tilde Q_a}.
\label{multicoset}
\end{equation}
Here $\tilde Q_a$ are those of the generators $\tilde Q_i$ that are spontaneously broken. We naturally assume that all the $Q_A$s are themselves spontaneously broken and that they are mutually linearly independent. This ensures that all the redundant fields $\x^\m_A$ can be eliminated by imposing a set of IHCs. With these assumptions, it is then straightforward to show that the nontrivial components of the MC form are
\begin{equation}
\begin{split}
\om_P^\m&=\dd x^\m,\\
\om^A_{K\m}&=(e^{-\im\t^a\Ka_a})^A_{\phantom AB}\dd\x^B_\m,\\
\om^A_Q&=(e^{-\im\t^a\Ka_a})^A_{\phantom AB}(\dd\t^B-\x^B\cdot\dd x),
\end{split}
\end{equation}
together with the MC form for the generators of $\mathfrak{g}$, which we denote as $\Omega$ as in eq.~\eqref{Omegaform}.

Next, we determine the transformation rules for all combinations of symmetries and fields. The symmetry transformations generated by $Q_A$ (with parameter $\a^A$) and $K^A_\m$ (with parameter $\b^A_\m$) take a very simple form
\begin{equation}
\t^A\to\t^A+\a^A+\b^A\cdot x,\qquad
\x^A_\m\to\x^A_\m+\b^A_\m.
\label{425}
\end{equation}
Thanks to the chosen parameterization~\eqref{multicoset} of the coset space, the NG fields $\t^a$ are left intact by these transformations. On the other hand, a symmetry transformation generated by $\tilde Q_i$, $e^{\im\eps^i\tilde Q_i}$, acts on these non-Galileon NG fields as in eq.~\eqref{gtransfo}. On the Galileon fields $\t^A$ and $\x^A_\m$, it acts linearly according to the representation $R$,
\begin{equation}
\t^A\to\bigl(e^{\im\eps^i\Ka_i}\bigr)^A_{\phantom AB}\t^B,\qquad
\x^A_\m\to\bigl(e^{\im\eps^i\Ka_i}\bigr)^A_{\phantom AB}\x^B_\m.
\label{426}
\end{equation}
The spacetime coordinate $x^\m$ is left intact by all these internal symmetry transformations.

Based on these symmetry transformation rules, we expect a very simple structure of the invariant action. Namely, the $\mathfrak{g}$-part of the MC form, $\Omega$, is completely independent of the Galileon fields $\t^A$ and $\x^A_\m$, and it only transforms nontrivially under the $\tilde Q_i$ generators themselves. Eq.~\eqref{426} tells us that $\t^A$ and $\x^A_\m$ transform linearly under the whole subalgebra $\mathfrak{g}$, both its unbroken and broken part. Under the transformations generated by $Q_A$ and $K_{\m A}$, these fields transform as a set of $n$ independent Galileon copies, see eq.~\eqref{425}. The Galileon and non-Galileon NG fields therefore to a large extent decouple.


\paragraph{Invariant contributions to the Lagrangian.} At the end of the day, the redundant modes $\x^A_\m$ have to be eliminated by imposing a set of IHCs,
\begin{equation}
\om^A_Q=0\quad\Rightarrow\quad
\x^A_\m=\de_\m\t^A.
\label{IHCgeneral}
\end{equation}
Since the vielbein stemming from $\om^\m_P$ is trivial, the only building blocks that we have at hand to construct invariant Lagrangians are then $\om^A_{K\m}=(e^{-\im\t^a\Ka_a})^A_{\phantom AB}\de_\m\de_\n\t^B\dd x^\n\equiv\om^A_{K\m\n}\dd x^\n$, and the MC form for the $\mathfrak{g}$-fields, $\Omega$. The latter can be used to build invariant operators using standard tensor methods; see ref.~\cite{Andersen:2014ywa} where all such invariant operators in spacetime dimension from one to four up to the fourth order of the derivative expansion have been classified. The form $\om^A_{K\m}$ can be used to build invariants in pretty much the same way, by taking a product of several factors of $\om^A_{K\m}$, or their covariant derivatives, and contracting the indices with invariant tensors of the unbroken symmetry group. The simplest example of such an invariant is $\delta_{AB}\om^A_{K\m\n}\om^{B\m\n}_K$. In this particular case, the factor $e^{-\im\t^a\Ka_a}$ drops out and thus such Lagrangians give no interactions between the Galileon and non-Galileon sectors. However, in case the representation $\Ka_i$ is reducible with respect to the unbroken part of $\mathfrak{g}$, one can easily construct more general operators of the type $c_{AB}\om^A_{K\m\n}\om^{B\m\n}_K$, where $c_{AB}$ is a symmetric invariant tensor of the unbroken subalgebra of $\mathfrak{g}$. In the extreme case that $\mathfrak{g}$ is completely broken, $c_{AB}$ can take arbitrary values. Another natural option how to generate interactions between the Galileon and the non-Galileon sector is to simply multiply invariant operators constructed separately from $\Omega$ and from $\om^A_{K\m}$.


\paragraph{WZ terms.} Since upon imposing the IHCs~\eqref{IHCgeneral}, $\om^A_{K\m}$ contains two derivatives of $\t^A$, canonical kinetic terms for the Galileon fields $\t^A$ can only be constructed as WZ terms, and it is therefore imperative to check whether such WZ terms exist. Following closely Witten's construction of WZ terms, used in ref.~\cite{Goon:2012dy} to obtain the Galileon Lagrangians in the single-flavor case, we search for these as invariant 5-forms that belong to the Lie algebra cohomology of the symmetry. To see how this works on a simple example, consider the 5-form
\begin{equation}
\om_5\equiv\eps_{\k\l\m\n}c_A\om^A_Q\wedge\dd x^\k\wedge\dd x^\l\wedge\dd x^\m\wedge\dd x^\n.
\end{equation}
Invariance of $\om_5$ under the internal symmetry requires invariance of $c_A$ under the \emph{unbroken} part of $\mathfrak{g}$, or more precisely under the representation $R$ thereof. Now recall the MC structure equation~\eqref{MCeq}, which in the present case implies that
\begin{equation}
\dd\om^A_Q=\dd x^\m\wedge\om^A_{K\m}-\im(\Ka_i)^A_{\phantom AB}\Omega^i\wedge\om^B_Q,\qquad
\dd\om^A_{K\m}=-\im(\Ka_i)^A_{\phantom AB}\Omega^i\wedge\om^B_{K\m}.
\end{equation}
The precise form of $\Omega^i$ depends on the pattern of symmetry breaking, but it must in any case contain a term linear in gradients of the NG fields $\t^a$, proportional to $\tilde Q_a\dd\t^a$. Hence for $\om_5$ to be closed, $c_A$ must be invariant under the \emph{broken} part of $\mathfrak{g}$ in the representation $R$, defined by the matrices $\Ka_i$. Altogether, invariance and closedness require that $c_A$ is invariant under the whole algebra $\mathfrak{g}$ in the representation $R$. It is then easy to see that $\om_5=\dd\om_4$, where
\begin{equation}
\om_4=\eps_{\k\l\m\n}c_A\tilde\t^A\dd x^\k\wedge\dd x^\l\wedge\dd x^\m\wedge\dd x^\n,
\end{equation}
where
\begin{equation}
\tilde\t^A\equiv(e^{-\im\t^a\Ka_a})^A_{\phantom AB}\t^B,\qquad
\tilde\x^A_\m\equiv(e^{-\im\t^a\Ka_a})^A_{\phantom AB}\x^B_\m.
\end{equation}
Using once more the required invariance of $c_A$ under the representation $R$ of $\mathfrak{g}$, this is easily seen to correspond to the tadpole Lagrangian density, $\La=c_A\t^A$.

Following this example and the analogy with the case of the single-flavor Galileon~\cite{Goon:2012dy}, we can now construct a whole class of WZ terms using as building blocks the 1-forms $\om_P^\m$, $\om^A_{K\m}$ and $\om^A_Q$,
\begin{equation}
\begin{split}
\om^1_5&=\eps_{\k\l\m\n}c_A\om^A_Q\wedge\dd x^\k\wedge\dd x^\l\wedge\dd x^\m\wedge\dd x^\n,\\
\om^2_5&=\eps_{\k\l\m\n}c_{AB}\om^A_Q\wedge\om^{B\k}_K\wedge\dd x^\l\wedge\dd x^\m\wedge\dd x^\n,\\
\om^3_5&=\eps_{\k\l\m\n}c_{ABC}\om^A_Q\wedge\om^{B\k}_K\wedge\om^{C\l}_K\wedge\dd x^\m\wedge\dd x^\n,\\
\om^4_5&=\eps_{\k\l\m\n}c_{ABCD}\om^A_Q\wedge\om^{B\k}_K\wedge\om^{C\l}_K\wedge\om^{D\m}_K\wedge\dd x^\n,\\
\om^5_5&=\eps_{\k\l\m\n}c_{ABCDE}\om^A_Q\wedge\om^{B\k}_K\wedge\om^{C\l}_K\wedge\om^{D\m}_K\wedge\om^{E\n}_K.
\end{split}
\end{equation}
Closedness and invariance under the internal symmetry require that the coefficients $c_{AB\dotsb}$ are invariant tensors of the representation $R$ of $\mathfrak{g}$, fully symmetric in all their coefficients. Explicit integration then shows that all these 5-forms belong to the Lie algebra cohomology, that is, are given by an exterior derivative of a noninvariant 4-form $\om_4$, where, in turn,
\begin{align}
\notag
\om^1_4&=\eps_{\k\l\m\n}c_A\tilde\t^A\dd x^\k\wedge\dd x^\l\wedge\dd x^\m\wedge\dd x^\n,\\
\notag
\om^2_4&=\eps_{\k\l\m\n}c_{AB}\bigl(\tilde\t^A\om^{B\k}_K\wedge\dd x^\l\wedge\dd x^\m\wedge\dd x^\n+\tfrac18\tilde\x^A\cdot\tilde\x^{B}\dd x^\k\wedge\dd x^\l\wedge\dd x^\m\wedge\dd x^\n\bigr),\\
\om^3_4&=\eps_{\k\l\m\n}c_{ABC}\bigl(\tilde\t^A\om^{B\k}_K\wedge\om^{C\l}_K\wedge\dd x^\m\wedge\dd x^\n+\tfrac13\tilde\x^A\cdot\tilde\x^{B}\om^{C\k}_K\wedge\dd x^\l\wedge\dd x^\m\wedge\dd x^\n\bigr),\\
\notag
\om^4_4&=\eps_{\k\l\m\n}c_{ABCD}\bigl(\tilde\t^A\om^{B\k}_K\wedge\om^{C\l}_K\wedge\om^{D\m}_K\wedge\dd x^\n+\tfrac34\tilde\x^A\cdot\tilde\x^{B}\om^{C\k}_K\wedge\om^{D\l}_K\wedge\dd x^\m\wedge\dd x^\n\bigr),\\
\notag
\om^5_4&=\eps_{\k\l\m\n}c_{ABCDE}\bigl(\tilde\t^A\om^{B\k}_K\wedge\om^{C\l}_K\wedge\om^{D\m}_K\wedge\om^{E\n}_K+2\tilde\x^A\cdot\tilde\x^{B}\om^{C\k}_K\wedge\om^{D\l}_K\wedge\om^{E\m}_K\wedge\dd x^\n\bigr).
\end{align}
The value of the coefficient in front of $\tilde\x^A\cdot\tilde\x^B$ in $\om^k_4$ agrees with the general value, valid in $D$ spacetime dimensions, $(k-1)/[2(D-k+2)]$, see e.g.~ref.~\cite{Novotny:2016jkh}.

Remarkably, all dependence on the non-Galileon NG fields $\t^a$ drops thanks to the required invariance of the coefficients $c_{AB\dotsb}$. What we get upon imposing the set of IHCs~\eqref{IHCgeneral} is the set of standard multi-Galileon Lagrangians. By introducing a shorthand notation for the antisymmetrized products of second derivatives of $\t^a$,
\begin{equation}
G^{A_1\dotsb A_k}_k\equiv\frac1{(4-k)!}\eps_{\a_1\dotsb\a_k\m_{k+1}\dotsb\m_4}\eps^{\b_1\dotsb\b_k\m_{k+1}\dotsb\m_4}(\de_{\b_1}\de^{\a_1}\t^{A_1})\dotsb(\de_{\b_k}\de^{\a_k}\t^{A_k}),
\end{equation}
where we set $G_0\equiv1$, and some manipulation using integration by parts, our multi-Galileon Lagrangians can be expressed as
\begin{equation}
\La_k=c_{A_1\dotsb A_k}\t^{A_1}G_{k-1}^{A_2\dotsb A_k}.
\label{multiGal}
\end{equation}
This is a direct generalization of the multi-Galileon Lagrangians discussed, for instance, in ref.~\cite{Goon:2012dy} to an arbitrary Lie algebra $\mathfrak{g}$ and its arbitrary real finite-dimensional representation $R$. The existence of such Lagrangians is only constrained by the existence of the fully symmetric invariant tensors $c_{AB\dotsb}$ for the given representation $R$.


\paragraph{Summary of invariant actions.} Altogether, we have found the following possible contributions to the action for the Galileon fields $\t^A$ and the non-Galileon NG fields $\t^a$. The former possess a set of WZ terms~\eqref{multiGal}, which provide their canonical kinetic terms and dominant interactions. While the existence of the WZ terms in general depends on the symmetry algebra, the kinetic term is generally present, since $c_{AB}=\delta_{AB}$ is an invariant tensor of any real representation of an arbitrary (compact) Lie algebra $\mathfrak{g}$.

The NG fields $\t^a$ likewise possess an infinite class of terms, independent of $\t^A$, which can be constructed using their MC form $\Omega^i\equiv\Omega^i_\m\dd x^\m$. The leading contribution to their Lagrangian is given by $c_{ab}\Omega^a_\m\Omega^{b\m}$, where $c_{ab}$ is a rank-two invariant tensor of the unbroken part of $\mathfrak{g}$~\cite{Andersen:2014ywa}. In higher orders of the derivative expansion, WZ terms may be present in this non-Galileon sector as well, and are known to be classified by the de Rham cohomology of the coset space of the broken symmetry~\cite{DHoker:1994ti,DHoker:1995it}.

Interaction terms bringing together the Galileon and non-Galileon fields may be easily constructed in higher orders of the derivative expansion. They are obtained either as invariant operators built out of the MC form $\om^A_{K\m}$, or as products thereof with invariants constructed out of $\Omega$.

Since the construction of strictly invariant contributions to the effective Lagrangian is a routine task once the MC form is known, let us conclude with some remarks regarding the WZ terms. While we have constructed the WZ terms~\eqref{multiGal} in analogy with the known multi-Galileon Lagrangians, we cannot exclude the existence of other WZ terms for the Galileon fields $\t^A$. Namely, all our WZ terms were obtained using the Levi-Civita tensor $\eps_{\k\l\m\n}$ to build a Lorentz-invariant 5-form. However, the Lorentz group has two additional invariant tensors of rank up to four, $g_{\m\n}$ and $g_{\k\l}g_{\m\n}$, which admit additional invariant 5-forms in case of several Galileon flavors such as
\begin{equation}
\begin{gathered}
c_{ABCD}\om_Q^A\wedge\om_Q^B\wedge\om_Q^C\wedge\om_{K\m}^D\wedge\dd x^\m,\qquad
c_{ABCD}\om_Q^A\wedge\om_{K\m}^B\wedge\om_{K\n}^C\wedge\om_K^{D\m}\wedge\dd x^\n,\\
c_{ABC}\om_Q^A\wedge\om_{K\m}^B\wedge\om_{K\n}^C\wedge\dd x^\m\wedge\dd x^\n,\qquad
c_{ABCDE}\om_Q^A\wedge\om_{K\m}^B\wedge\om_{K\n}^C\wedge\om_K^{D\m}\wedge\om_K^{E\n}.
\end{gathered}
\end{equation}
We have not performed an exhaustive search here, but we have checked that the two 5-forms on the first line above, upon integration and imposing the IHCs~\eqref{IHCgeneral}, lead to Lagrangian densities that are total derivatives, and hence do not affect the perturbative physics of the NG modes.

Likewise, we cannot on general grounds exclude the existence of WZ terms that mix the Galileon and non-Galileon fields in that they are constructed out of \emph{both} $\om^A_Q$ and $\Omega$. We however expect that, if possible at all, such terms will be strongly constrained by symmetry, as opposed to the general WZ terms~\eqref{multiGal} that exist for an infinite class of Lie algebras and their representations.


\subsubsection{Generalized DBI solutions}

The generalized Galileon solutions are characterized by vanishing $f_{AB}$. We will now show that a similar construction can be carried out in the opposite limit, that is when $f_{AB}$ is assumed to be nonsingular.

With this assumption, we can use $f_{AB}$ and its inverse as a metric to raise and lower indices. Using the experience gained in the analysis of the single-flavor case in section~\ref{subsec:DBI}, we can now redefine the generators $Q_i$ as
\begin{equation}
\tilde Q_i\equiv Q_i+h_{Ai}f^{AB}Q_B.
\end{equation}
It is then a matter of straightforward algebra using the Jacobi constraints~\eqref{eqai}--\eqref{eqfxi} to show that the commutation relations including the generators $Q_i$ and $K_{\m A}$ reduce to
\begin{align}
\notag
[P_\m,K_{\n A}]&=\im g_{\m\n}Q_A,\\
\notag
[K_{\m A},K_{\n B}]&=\im(f_{AB}J_{\m\n}+g_{\m\n}Q_{AB}),\\
\notag
[\tilde Q_i,K_{\m A}]&=(\Ka_i)^B_{\phantom BA}K_{\m B},\\
\label{DBIalgebra}
[K_{\m A},Q_B]&=-\im f_{AB}P_\m,\\
\notag
[Q_A,Q_B]&=0,\\
\notag
[\tilde Q_i,Q_A]&=(\Ka_i)^B_{\phantom BA}Q_B,\\
\notag
[\tilde Q_i,\tilde Q_j]&=\im\str^k_{ij}\tilde Q_k,
\end{align}
where we additionally defined $Q_{AB}\equiv\hh^i_{AB}Q_i$. Since $\tilde Q_A\equiv a^i_A\tilde Q_i=0$, the set of generators $Q_i$ splits up into the $Q_A$s and those of the $\tilde Q_i$s that are nonzero. Accordingly, the Lie algebra of these internal symmetry generators acquires the structure of a semidirect sum of the subalgebra generated by the $\tilde Q_i$s and the Abelian subalgebra generated by the $Q_A$s. The former acts on the latter through the representation $\Ka_i$.

At this point, we can forget about all the constraints~\eqref{eqai}--\eqref{eqfxi}, for the commutation relations are fully determined by the Lie algebra of the generators $\tilde Q_i$, its representation $\Ka_i$ and the metric $f_{AB}$. Moreover, eqs.~\eqref{eqai}, \eqref{eqail} and \eqref{eqih} together with $f_{AB}=-a^i_Ah_{Bi}$ imply that $f_{AB}$ is an invariant metric of the representation $t_i$, that is,
\begin{equation}
(\Ka_i)^C_{\phantom CA}f_{CB}+(\Ka_i)^C_{\phantom CB}f_{AC}=0.
\end{equation}
The structure defined by eq.~\eqref{DBIalgebra} generalizes the set of commutators in eq.~\eqref{commuv} with $u=0$ to multiple flavors of the shift generators $Q_A$, and we will therefore refer to it as the generalized DBI theory. It has an elegant geometric interpretation~\cite{Bogers:2018kuw}. Note that the linear combinations $Q_{AB}$ satisfy the commutation relations
\begin{equation}
\begin{split}
[Q_{AB},Q_{CD}]&=\im(f_{AD}Q_{BC}+f_{BC}Q_{AD}-f_{AC}Q_{BD}-f_{BD}Q_{AC}),\\
[Q_{AB},Q_C]&=\im(f_{BC}Q_A-f_{AC}Q_B),\\
[Q_{AB},K_{\m C}]&=\im(f_{BC}K_{\m A}-f_{AC}K_{\m B}).
\end{split}
\end{equation}
These together with the other commutation relations, listed above, imply that $J_{\m\n}$, $K_{\mu A}$, $Q_{AB}$, $P_\m$ and $Q_A$ generate a group of isometries of an extended spacetime with the metric $g_{\m\n}\oplus f_{AB}$. The generators $Q_{AB}$ play the role of rotations in the extra dimensions, labeled by $A,B,\dotsc$, $Q_A$ that of translations therein, and finally $K_{\mu A}$ that of rotations between the physical (Minkowski) and the extra dimensions. The remaining internal generators $\tilde Q_i$ act on the extra-dimensional coordinates through the representation $\Ka_i$. The generators $Q_{AB}$ then naturally form a rank-two antisymmetric tensor under this representation,
\begin{equation}
[\tilde Q_i,Q_{AB}]=(\Ka_i)^C_{\phantom CA}Q_{CB}+(\Ka_i)^C_{\phantom CB}Q_{AC}.
\end{equation}


\paragraph{Coset construction.} In order to proceed, we use the same coset parameterization as in the generalized Galileon case, eq.~\eqref{multicoset}. It is then straightforward to evaluate \emph{some} of the components of the MC form without making further simplifying assumptions. Using the shorthand notation
\begin{equation}
\Co(x)\equiv\cosh\sqrt x,\quad
\Si(x)\equiv\frac{\sinh\sqrt x}{\sqrt x},\quad
\XX_\m^{\phantom\m\n}\equiv f_{AB}\x^A_\m\x^{\n B},\quad
\LL_A^{\phantom AB}\equiv f_{AC}\x^{\m C}\x_\m^B,
\end{equation}
we obtain
\begin{equation}
\begin{split}
\om^\m_P&=\dd x^\n(\Co\XX)_\n^{\phantom\n\m}-\dd\t^Af_{AB}\x^{\n B}(\Si\XX)_\n^{\phantom\n\m},\\
\om^A_Q&=(e^{-\im\t^a\Ka_a})^A_{\phantom AB}\bigl[\dd\t^C(\Co\LL)_C^{\phantom CB}-\dd x^\m\x_\m^C(\Si\LL)_C^{\phantom CB}\bigr].
\end{split}
\label{DBIMC}
\end{equation}
The coset construction also gives us the symmetry transformation rules. The extended spacetime translations act as expected and amount to trivial shifts of the coordinate $x^\m$ (by the generator $P_\m$) and of $\t^A$ (by the generator $Q_A$), respectively. The transformations generated by $\tilde Q_i$ act linearly as in eq.~\eqref{426}. The transformations generated by $K^A_\m$, with parameter $\b^A_\m$, take a more complicated form this time,
\begin{equation}
\begin{split}
x^\m&\to x^\n(\Co\XX_\b)_\n^{\phantom\n\m}+\t^Af_{AB}\b^{\n B}(\Si\XX_\b)_\n^{\phantom\n\m},\\
\t^A&\to\t^B(\Co\LL_\b)_B^{\phantom BA}+x^\m\b^B_\m(\Si\LL_\b)_B^{\phantom BA},\\
\x^A_\m&\to\x^A_\m+\b^A_\m+\mathcal O(\b^2,\x^2),
\end{split}
\end{equation}
where $(\XX_\b)_\m^{\phantom\m\n}\equiv f_{AB}\b^A_\m\b^{\n B}$ and $(\LL_\b)_A^{\phantom AB}\equiv f_{AC}\b^{\m C}\b^B_\m$.


\paragraph{Invariant actions.} In analogy with the single-flavor case, we do not expect any interesting WZ terms for the generalized DBI theory. We can therefore focus on the construction of strictly invariant Lagrangians.

At the end of the day, the redundant modes $\x^A_\m$ are disposed of by imposing a set of IHCs, $\om^A_Q=0$. This gives $\x^A_\m$ implicitly in terms of $\de_\m\t^A$ as a solution of the condition
\begin{equation}
\de_\m\t^A=\x^B_\m\left(\frac{\Si\LL}{\Co\LL}\right)_B^{\phantom BA}.
\label{DBIIHC}
\end{equation}
The leading-order action is then given solely by integrating the invariant volume measure, $\dd^4x\sqrt{-G}$. The metric $G_{\m\n}\equiv g_{\a\b}e^\a_\m e^\b_\n$ is in turn constructed from the vielbein, extracted from $\om^\m_P$. Using the IHC~\eqref{DBIIHC} leads to
\begin{equation}
G_{\m\n}=g_{\m\n}-f_{AB}\de_\m\t^A\de_\n\t^B,
\end{equation}
as a direct multi-flavor generalization of eq.~\eqref{metric}.

Similarly to the generalized Galileon case, the leading-order action built up from this metric is independent of the non-DBI NG fields $\t^a$ altogether. Yet, just as in the generalized Galileon case, the form of the action is a nontrivial result. The metric $f_{AB}$ appearing here (just like the coefficients $c_{AB\dotsb}$ in the Galileon case) is namely constrained to be invariant under the \emph{whole} algebra of $\tilde Q_i$, regardless of which of its generators are spontaneously broken and which remain unbroken.

The leading-order action for the non-DBI fields $\t^a$, and at the same time the leading interaction between these and the DBI fields $\t^A$, is obtained using the MC form $\Omega^i$, and takes the form
\begin{equation}
\int\dd^4 x\sqrt{-G}\,c_{ab}\Omega^a_\m\Omega^{b\m},
\end{equation}
where $c_{ab}$ is a rank-two invariant tensor of the unbroken part of the algebra of $\tilde Q_i$. Higher-order actions can be likewise constructed by putting together more factors of $\Omega^a_\m$, by employing $\om^{\m A}_K$, or by taking their covariant derivatives. These, however, require the knowledge of the spin connection, not evaluated here.


\section{Summary and conclusions}
\label{sec:conclusions}

In this paper and its companion~\cite{Bogers:2018kuw}, we have initiated a classification of effective theories featuring enhanced soft limits from the symmetry point of view. The motivation for this work was the fact that physical massless scalars are always NG bosons of a spontaneously broken global symmetry. Our main tool was the Lie-algebraic classification of extensions of the physical symmetry by adding a set of additional, spontaneously broken but redundant, generators.

To warm up, we analyzed Lorentz-invariant theories for a single physical NG boson. As expected, we only ``rediscovered'' theories that are already well known. Our approach, however, helped to clarify the relation between the Galileon and DBI theories, and to shed new light on the extended symmetry of the special Galileon. Next, we focused on Lorentz-invariant theories featuring several physical NG bosons. With some simplifying assumptions on the symmetry Lie algebra, we then found two infinite classes of algebraic structures, leading to effective theories combining NG bosons with and without enhanced soft limits. These classes contain as their special cases all the known theories of the multi-Galileon and multi-flavor DBI type. Concrete theories in these classes are determined by choosing a set of geometric data such as a Lie algebra and its real finite-dimensional representation, or (in the DBI case) an invariant metric on the target space of the representation. A fully general solution to all the Lie-algebraic constraints on symmetries possessing redundant generators is given in Ref.~\cite{Bogers:2018kuw}.

In our future work, we plan to analyze in detail some concrete examples of the multi-flavor theories constructed here. More importantly, however, we will extend the framework developed here to systems lacking Lorentz invariance, commonly found in condensed-matter physics. In the context of high-energy physics, one may think of redundant symmetries as a useful tool to generate actions for NG bosons with enhanced scattering amplitudes. In condensed-matter physics, on the contrary, there are numerous examples of naturally occurring physical symmetries that are redundant, for instance Galilei boosts in superfluids or spatial rotations in crystals~\cite{Watanabe:2013iia,Brauner:2014aha,Radzihovsky:2011ra,Beekman:2013na}. While the problem of mere counting of NG bosons in such systems is by now well understood, we plan to initiate their study from the point of view of scaling of scattering amplitudes.

It would also be interesting to clarify to what extent the framework developed here can be used to study the scattering amplitudes of higher-spin massless particles. Recent studies showed that some of the scalar theories discussed here can be recovered through dimensional reduction of higher-dimensional theories of spin-one and spin-two particles~\cite{Cheung:2017ems,Cheung:2017yef}. On the other hand, the investigation of vector effective field theories with enhanced soft limits of scattering amplitudes was initiated in ref.~\cite{Cheung:2018oki}. Nevertheless, it remains to be seen whether such exceptional theories can be addressed from the symmetry point of view, which is the starting assumption of our approach.


\acknowledgments
We are grateful to Karol Kampf, Ji\v{r}\'{\i} Novotn\'{y} and Jaroslav Trnka for discussions of details of their work at various stages of this project. This work has been supported by the grant no.~PR-10614 within the ToppForsk-UiS program of  the University of Stavanger and the University Fund.


\appendix


\section{Summary of the results}
\label{app:summary}

In this appendix, we give an overview of systems featuring enhanced soft limits of scattering amplitudes that were discussed throughout this paper, without the clutter of the intermediate steps of all their derivations. In each case, we list merely the corresponding Lie algebra and the basic building blocks for the construction of invariant actions. This is meant to allow others to use our results without having to go through the technical details.


\subsection{Single NG boson and singly enhanced soft limit}
\label{app:A1}

\subsubsection*{Lie algebra}

The symmetry generators include those of spacetime rotations ($J_{\m\n}$), spacetime translations ($P_\m$) and the spontaneously broken internal symmetry ($Q$), and a vector of redundant generators ($K_\m$). The nontrivial commutation relations of the Lie algebra read
\begin{align}
\notag
[J_{\m\n},J_{\k\l}]&=\im(g_{\m\l}J_{\n\k}+g_{\n\k}J_{\m\l}-g_{\m\k}J_{\n\l}-g_{\n\l}J_{\m\k}),\\
\notag
[J_{\m\n},P_\l]&=\im(g_{\n\l}P_\m-g_{\m\l}P_\n),\\
[J_{\m\n},K_\l]&=\im(g_{\n\l}K_\m-g_{\m\l}K_\n),\\
\notag
[P_\m,K_\n]&=\im g_{\m\n}Q,\\
\notag
[K_\m,K_\n]&=-\im vJ_{\m\n},\\
\notag
[K_\m,Q]&=\im vP_\m,
\end{align}
where $v$ is a real parameter. This general structure includes the Galileon algebra ($v=0$) and the five-dimensional Poincar\'e algebra ($v\neq0$), leading to the DBI theory. Note that there is a further extension of this algebra, discussed in section~\ref{subsec:simple}. It is isomorphic to the five-dimensional conformal algebra, but does not lead to enhanced soft limits, and thus is omitted from the overview given here.


\subsection*{Building blocks for invariant actions}

The basic building blocks for invariant actions stem from the MC form, and are given by:\\[-3ex]
\begin{itemize}
\itemsep0pt
\item The vielbein,
\begin{equation}
e^\a_\m=\delta^\a_\m+\frac{\x^\a\x_\m}{\x^2}\left(\frac1{\cos\sqrt{v\x^2}}-1\right),
\end{equation}
where the auxiliary field $\x^\m$ is defined implicitly by
\begin{equation}
\de_\m\t=\x_\m\frac{\tan\sqrt{v\x^2}}{\sqrt{v\x^2}},
\end{equation}
and $\t$ is the physical NG boson field.
\item  The metric, induced by the vielbein,
\begin{equation}
G_{\m\n}=g_{\a\b}e^\a_\m e^\b_\n=g_{\m\n}+v\de_\m\t\de_\n\t.
\end{equation}
\item Covariant derivative of the auxiliary field,
\begin{equation}
\nabla_\m\x^\n=\left[\delta^\n_\a\frac{\sin\sqrt{v\x^2}}{\sqrt{v\x^2}}+\frac{\x^\n\x_\a}{\x^2}\left(\cos\sqrt{v\x^2}-\frac{\sin\sqrt{v\x^2}}{\sqrt{v\x^2}}\right)\right]\de_\m\x^\a.
\end{equation}
\end{itemize}
If needed, higher-order covariant derivatives, or in general covariant derivatives of tensor fields, are accomplished using the spin connection,
\begin{equation}
\om^{\m\n}_\l=\frac1{\x^2}(1-\cos\sqrt{v\x^2})(\x^\n\de_\l\x^\m-\x^\m\de_\l\x^\n).
\end{equation}
The invariant volume measure in the action then reads $\dd^4x\sqrt{-G}=\dd^4x\sqrt{1+v\de_\m\t\de^\m\t}$, and invariant Lagrangian densities are constructed from products of tensor fields with their indices contracted by the metric $G_{\m\n}$ or its inverse.\footnote{Note that the metric $G_{\m\n}$ and its inverse $G^{\m\n}$ can likewise be used to lower and raise indices of all the covariant objects listed here. However, the Lorentz indices appearing \emph{inside} their definitions, for instance those on $\x^\a\x_\m$ in $e^\a_\m$, are naturally raised and lowered using the flat-space Minkowski metric $g_{\m\n}$. The same remark applies to all the other theories listed in this appendix.}

In addition to strictly invariant Lagrangian densities, the symmetry algebra admits a set of WZ terms in the $v=0$ (Galileon) case. These coincide with the standard Galileon terms, see ref.~\cite{Goon:2012dy} for more details. It is also shown therein that the more general $v\neq0$ (DBI) case admits a single WZ term, corresponding to $\La=\t$, the tadpole Lagrangian. This should, however, be omitted from any consistent theory of an interacting massless scalar. All physically relevant Lagrangians can in the DBI case therefore be obtained using the procedure outlined above.


\subsection{Single NG boson and doubly enhanced soft limit}

\subsubsection*{Lie algebra}

In addition to the generators $J_{\m\n}$, $P_\m$, $Q$ and $K_\m$, the Lie algebra now contains an additional traceless symmetric tensor $S^{\m\n}$. The nontrivial commutators of the Lie algebra read
\begin{equation}
\begin{split}
[J_{\m\n},J_{\k\l}]&=\im(g_{\m\l}J_{\n\k}+g_{\n\k}J_{\m\l}-g_{\m\k}J_{\n\l}-g_{\n\l}J_{\m\k}),\\
[J_{\m\n},P_\l]&=\im(g_{\n\l}P_\m-g_{\m\l}P_\n),\\
[J_{\m\n},K_\l]&=\im(g_{\n\l}K_\m-g_{\m\l}K_\n),\\
[J_{\m\n},S_{\k\l}]&=\im(-g_{\m\l}S_{\n\k}+g_{\n\k}S_{\m\l}-g_{\m\k}S_{\n\l}+g_{\n\l}S_{\m\k}),\\
[P_\m,K_\n]&=\im g_{\m\n}Q,\\
[S_{\m\n},S_{\k\l}]&=\im s(g_{\m\l}J_{\n\k}+g_{\n\k}J_{\m\l}+g_{\m\k}J_{\n\l}+g_{\n\l}J_{\m\k}),\\
[S_{\m\n},P_\l]&=\im(g_{\m\l}K_\n+g_{\n\l}K_\m-\tfrac12g_{\m\n} K_\l),\\
[S_{\m\n},K_\l]&=\im s(g_{\m\l}P_\n+g_{\n\l}P_\m-\tfrac12g_{\m\n}P_\l),
\end{split}
\end{equation}
where $s=\pm1$. (The case $s=0$, also discussed in section~\ref{subsec:double}, does not lead to any nontrivial models with a doubly enhanced soft limit.) Both signs correspond to the symmetry of the special Galileon. Note that it is also possible to extend the Lie algebra of $J_{\m\n}$, $P_\m$, $Q$, $K_\m$ by adding a Lorentz-singlet redundant generator. This, however, does not lead to any theories of an interacting NG boson with a nontrivially realized doubly enhanced soft limit.


\subsection*{Building blocks for invariant actions}

The basic building blocks for invariant actions, as given by the MC form, are:\\[-3ex]
\begin{itemize}
\itemsep0pt
\item The vielbein,
\begin{equation}
e^\a_\m=\cosh(\sqrt s\b)^\a_{\phantom\a\m}+\sqrt s\,\de_\m\x_\n\sinh(\sqrt s\b)^{\a\n},
\end{equation}
where $\x^\m$ is an auxiliary vector and $\b^{\m\n}$ is an auxiliary traceless symmetric tensor.
\item The metric, induced by the vielbein, $G_{\m\n}=g_{\a\b}e^\a_\m e^\b_\n$.
\item Covariant derivative of the auxiliary vector field, $\nabla_\m\x^\n$, defined implicitly by
\begin{equation}
e^\a_\n\nabla_\m\x^\n=\frac{\sinh(\sqrt s\b)^\a_{\phantom\a\m}}{\sqrt s}+\de_\m\x_\n\cosh(\sqrt s\b)^{\a\n}.
\end{equation}
\item Covariant derivative of the auxiliary tensor field, $\nabla_\l\b^{\m\n}$, defined implicitly by
\begin{equation}
e^\a_\m e^\b_\n\nabla_\l\b^{\m\n}=\de_\l\b^{\m\n}\frac{\bigl[B^{-1}\sinh(\sqrt sB)\bigr]^{\a\b}_{\m\n}}{\sqrt s},
\end{equation}
where
\begin{equation}
B^{\a\b}_{\m\n}\equiv\b^\a_\m\delta^\b_\n-\b^\b_\n\delta^\a_\m.
\end{equation}
\end{itemize}
The auxiliary vector field is eliminated by setting $\x_\m=\de_\m\t$, where $\t$ is the physical NG boson field. The auxiliary tensor field is eliminated by setting the traceless symmetric part of $\nabla_\m\x_\n$ to zero. The remaining building blocks for the construction of invariant actions are then the singlet covariant derivative $\nabla_\m\x^\m$, the antisymmetric part of $\nabla_\m\x_\n$, and $\nabla_\l\b^{\m\n}$. Covariant derivatives of these building blocks can be constructed using the spin connection,
\begin{equation}
\om^{\m\n}_\l=\de_\l\b^{\a\b}\bigl\{B^{-1}[\cosh(\sqrt sB)-\mathbbm1]\bigr\}^{\m\n}_{\a\b}.
\end{equation}
In four spacetime dimensions, there is a single WZ term admitted by the symmetry algebra, corresponding to the special Galileon; see appendix~\ref{app:WZterms} for details.


\subsection{Multiple NG bosons: DBI-like theory with a single redundant generator}

\subsubsection*{Lie algebra}

In this case, the Poincar\'e algebra is extended by an arbitrary set of internal symmetry generators ($Q_i$) and a single redundant vector ($K_\m$). In the particular solution of all the Lie-algebraic constraints, generalizing the DBI theory, the generators $Q_i$ can be split up into $\tilde Q$ and $\tilde Q_i$ such that
\begin{align}
\notag
[J_{\m\n},J_{\k\l}]&=\im(g_{\m\l}J_{\n\k}+g_{\n\k}J_{\m\l}-g_{\m\k}J_{\n\l}-g_{\n\l}J_{\m\k}),\\
\notag
[J_{\m\n},P_\l]&=\im(g_{\n\l}P_\m-g_{\m\l}P_\n),\\
\notag
[J_{\m\n},K_\l]&=\im(g_{\n\l}K_\m-g_{\m\l}K_\n),\\
\notag
[P_\m,K_\n]&=\im g_{\m\n}\tilde Q,\\
[K_\m,K_\n]&=-\im vJ_{\m\n},\\
\notag
[K_\m,\tilde Q]&=\im vP_\m,\\
\notag
[K_\m,\tilde Q_i]&=0,\\
\notag
[\tilde Q,\tilde Q_i]&=0,\\
\notag
[\tilde Q_i,\tilde Q_j]&=\im\str^k_{ij}\tilde Q_k.
\end{align}
In this basis, the Lie algebra is manifestly given by a direct sum of that of the DBI theory, reviewed in appendix~\ref{app:A1}, and that of the internal symmetry generators $\tilde Q_i$.


\subsection*{Building blocks for invariant actions}

The construction of invariant actions for the DBI part of the theory is explained in appendix~\ref{app:A1}. The non-Abelian sector of the generators $\tilde Q_i$ contributes through its own MC form $\Omega$, defined by
\begin{equation}
\Omega_\m\equiv-\im e^{-\im\tilde\t^a\tilde Q_a}\de_\m e^{\im\tilde\t^a\tilde Q_a},
\end{equation}
where $\tilde\t^a$ are the NG fields, associated with the broken generators $\tilde Q_a$. The broken part of $\Omega_\m$ represents the covariant derivative of the physical NG fields $\t^a$. The unbroken part, on the other hand, is needed to construct covariant derivatives of fields that transform nontrivially under the internal symmetry.


\subsection{Multiple NG bosons: general multi-Galileon theory}

\subsubsection*{Lie algebra}

This is a different extension of the Poincar\'e algebra, allowing for multiple redundant vector generators, labeled as $K^A_\m$. The nontrivial commutation relations of the algebra read
\begin{equation}
\begin{split}
[J_{\m\n},J_{\k\l}]&=\im(g_{\m\l}J_{\n\k}+g_{\n\k}J_{\m\l}-g_{\m\k}J_{\n\l}-g_{\n\l}J_{\m\k}),\\
[J_{\m\n},P_\l]&=\im(g_{\n\l}P_\m-g_{\m\l}P_\n),\\
[J_{\m\n},K_{\l A}]&=\im(g_{\n\l}K_{\m A}-g_{\m\l}K_{\n A}),\\
[P_\m,K_{\n A}]&=\im g_{\m\n}Q_A,\\
[K_{\m A},K_{\n B}]&=0,\\
[Q_i,K_{\m A}]&=(\Ka_i)^B_{\phantom BA}K_{\m B},\\
[Q_i,Q_j]&=\im\str^k_{ij}Q_k,\\
[Q_i,Q_A]&=(\Ka_i)^B_{\phantom BA}Q_B,\\
[Q_A,Q_B]&=0,
\end{split}
\end{equation}
where $Q_A\equiv a^i_AQ_i$ is a set of particular linear combinations of the generators $Q_i$, corresponding to those NG modes whose scattering amplitudes feature enhanced soft limits. Furthermore, $(\Ka_i)^B_{\phantom BA}$ is a set of matrices that define a representation of the internal symmetry group generated by $Q_i$. Below, we construct a class of theories of this type, where all the generators $Q_i$ can be split into two sets, $\tilde Q_i$ and $Q_A$, both of which form a closed Lie algebra. Any theory in this class is determined by specifying the Lie algebra $\mathfrak{g}$ of the generators $\tilde Q_i$ and its representation $R$, defined by the matrices $(\Ka_i)^B_{\phantom BA}$. The full internal symmetry algebra is isomorphic to the semidirect sum $\mathfrak{g}\ltimes\mathbb{R}^n$, $n$ being the dimension of $R$.


\subsection*{Building blocks for invariant actions}

Invariant actions are constructed as functionals of a set of Galileon fields $\t^A$, one for each generator $Q_A$, and the NG fields $\t^a$, one for each spontaneously broken generator $\tilde Q_a$. The basic building blocks are:\\[-3ex]
\begin{itemize}
\itemsep0pt
\item The vielbein, which is trivial for this class of theories, $e^\a_\m=\delta^\a_\m$.
\item Covariant derivative of the auxiliary field $\x^\m_A$,
\begin{equation}
\nabla_\m\x^{\n A}=(e^{-\im\t^a\Ka_a})^A_{\phantom AB}\de_\m\x^{B\n},
\end{equation}
which is related to $\t^A$ by $\x^A_\m=\de_\m\t^A$.
\item The MC form for the generators of $\mathfrak{g}$,
\begin{equation}
\Omega_\m\equiv-\im e^{-\im\t^a\tilde Q_a}\de_\m e^{\im\t^a\tilde Q_a}.
\end{equation}
The broken part thereof, $\Omega^a_\m$, represents the covariant derivative of the physical NG fields, $\nabla_\m\t^a$. The unbroken part, on the other hand, is needed to construct covariant derivatives of fields transforming nontrivially under the internal symmetry algebra $\mathfrak{g}$.
\end{itemize}
In addition to strictly invariant Lagrangians, built out of $\nabla_\m\x^{A\n}$ and $\Omega_\m^a$ and their covariant derivatives, there is a series of WZ terms, expressed compactly as
\begin{equation}
\La_k=c_{A_1\dotsb A_k}\t^{A_1}G_{k-1}^{A_2\dotsb A_k},
\end{equation}
where
\begin{equation}
G^{A_1\dotsb A_k}_k\equiv\frac1{(4-k)!}\eps_{\a_1\dotsb\a_k\m_{k+1}\dotsb\m_4}\eps^{\b_1\dotsb\b_k\m_{k+1}\dotsb\m_4}(\de_{\b_1}\de^{\a_1}\t^{A_1})\dotsb(\de_{\b_k}\de^{\a_k}\t^{A_k})
\end{equation}
for $k=1,\dotsc,4$, and $G_0\equiv1$. Here $c_{AB\dotsb}$ are required to be fully symmetric invariant tensors of $\mathfrak{g}$ in the representation $R$.


\subsection{Multiple NG bosons: general multi-flavor DBI theory}

\subsubsection*{Lie algebra}

This extension of the Poincar\'e algebra again contains multiple redundant vector generators, $K^A_\m$. The internal generators $Q_i$ can be divided into subsets $Q_A$ and $\tilde Q_i$ such that
\begin{equation}
\begin{split}
[J_{\m\n},J_{\k\l}]&=\im(g_{\m\l}J_{\n\k}+g_{\n\k}J_{\m\l}-g_{\m\k}J_{\n\l}-g_{\n\l}J_{\m\k}),\\
[J_{\m\n},P_\l]&=\im(g_{\n\l}P_\m-g_{\m\l}P_\n),\\
[J_{\m\n},K_{\l A}]&=\im(g_{\n\l}K_{\m A}-g_{\m\l}K_{\n A}),\\
[P_\m,K_{\n A}]&=\im g_{\m\n}Q_A,\\
[K_{\m A},K_{\n B}]&=\im(f_{AB}J_{\m\n}+g_{\m\n}Q_{AB}),\\
[K_{\m A},Q_B]&=-\im f_{AB}P_\m,\\
[\tilde Q_i,K_{\m A}]&=(\Ka_i)^B_{\phantom BA}K_{\m B},\\
[\tilde Q_i,\tilde Q_j]&=\im\str^k_{ij}Q_k,\\
[\tilde Q_i,Q_A]&=(\Ka_i)^B_{\phantom BA}Q_B,\\
[Q_A,Q_B]&=0,
\end{split}
\end{equation}
where the generators $Q_A$ correspond to the NG modes featuring scattering amplitudes with enhanced soft limits. The matrices $(\Ka_i)^B_{\phantom BA}$ define a representation of the internal symmetry subalgebra with generators $\tilde Q_i$. The matrix of coefficients $f_{AB}$ is a rank-two symmetric invariant tensor of this representation. The generators $Q_{AB}$, appearing on the right-hand side of $[K_{\m A},K_{\n B}]$, satisfy the relations
\begin{equation}
\begin{split}
[Q_{AB},Q_{CD}]&=\im(f_{AD}Q_{BC}+f_{BC}Q_{AD}-f_{AC}Q_{BD}-f_{BD}Q_{AC}),\\
[Q_{AB},Q_C]&=\im(f_{BC}Q_A-f_{AC}Q_B),\\
[Q_{AB},K_{\m C}]&=\im(f_{BC}K_{\m A}-f_{AC}K_{\m B}),\\
[\tilde Q_i,Q_{AB}]&=(\Ka_i)^C_{\phantom CA}Q_{CB}+(\Ka_i)^C_{\phantom CB}Q_{AC}.
\end{split}
\end{equation}
Altogether, the generators $J_{\m\n}$, $K_{\m A}$, $Q_{AB}$, $P_\m$ and $Q_A$ span the Lie algebra of infinitesimal isometries of the extended spacetime with the metric $g_{\m\n}\oplus f_{AB}$. Any theory in this class is determined by specifying the Lie algebra of the generators $\tilde Q_i$, its representation $\Ka_i$ and the invariant metric $f_{AB}$.


\subsection*{Building blocks for invariant actions}

The physical degrees of freedom are the DBI fields $\t^A$, one for each generator $Q_A$, and the non-DBI NG fields $t^a$, one for each spontaneously broken generator $\tilde Q_a$. Invariant actions are then constructed using the following building blocks:\\[-3ex]
\begin{itemize}
\itemsep0pt
\item The vielbein,
\begin{equation}
e^\a_\m=(\Co\XX)_\m^{\phantom\m\a}-\de_\m\t^Af_{AB}\x^{\n B}(\Si\XX)_\n^{\phantom\n\a},
\end{equation}
where
\begin{equation}
\Co(x)\equiv\cosh\sqrt x,\quad
\Si(x)\equiv\frac{\sinh\sqrt x}{\sqrt x},\quad
\XX_\m^{\phantom\m\n}\equiv f_{AB}\x^A_\m\x^{\n B},\quad
\LL_A^{\phantom AB}\equiv f_{AC}\x^{\m C}\x_\m^B,
\end{equation}
and the auxiliary vector field $\x^\m_A$ is defined implicitly through
\begin{equation}
\de_\m\t^A=\x^B_\m\left(\frac{\Si\LL}{\Co\LL}\right)_B^{\phantom BA}.
\end{equation}
\item The metric, induced by the vielbein,
\begin{equation}
G_{\m\n}=g_{\a\b}e^\a_\m e^\b_\n=g_{\m\n}-f_{AB}\de_\m\t^A\de_\n\t^B.
\end{equation}
\item Covariant derivative of the auxiliary field, $\nabla_\m\x^{\n A}$, not evaluated here.
\item The MC form for the generators $\tilde Q_i$,
\begin{equation}
\Omega_\m\equiv-\im e^{-\im\t^a\tilde Q_a}\de_\m e^{\im\t^a\tilde Q_a}.
\end{equation}
Its broken part, $\Omega^a_\m$, represents the covariant derivative of the physical NG fields, $\nabla_\m\t^a$. The unbroken part, on the other hand, is needed to construct covariant derivatives of fields transforming nontrivially under the algebra of $\tilde Q_i$.
\end{itemize}
Covariant derivatives of tensor fields are obtained using the spin connection, not evaluated here. The invariant volume measure for the action reads $\dd^4x\sqrt{-G}$. Invariant Lagrangian densities are constructed from products of tensor fields with their indices contracted by the metric $G_{\m\n}$ or its inverse.


\section{Choosing the basis of the Lie algebra}
\label{app:basis}

When classifying possible Lie-algebraic structures associated with redundant symmetries, it is important to take into account the freedom to choose a basis of the Lie algebra; we saw in section~\ref{subsec:physicalsol} that even apparently quite different commutation relations can in fact correspond to the same Lie algebra. Apart from a trivial rescaling of some of the generators which allows us to eliminate some of the free parameters in the commutation relations, we often encounter the situation that a generator $X$ acts on a two-dimensional subspace spanned on two other generators, $A$ and $B$, as a linear mapping,
\begin{equation}
[X,A]=\im(aA+bB),\qquad
[X,B]=\im(cA+dB).
\end{equation}
The matrix of coefficients $a,b,c,d$ can be reduced by changing the basis using the following elementary statement from linear algebra, which we formulate as a simple theorem:
\begin{theorem}
\label{thm:22matrix}
Every real $2\times2$ matrix $M$ can by a real similarity transformation be brought to the form
\begin{equation}
\biggl(\begin{matrix}
\k & \l\\
s\l & \k
\end{matrix}\biggr),
\label{theorem}
\end{equation}
where $\kappa=\frac12\mathrm{tr}M$ and $s=\sgn\bigl[(\mathrm{tr}M)^2-4\det M\bigr]$, and $\lambda$ is real and non-negative.
\end{theorem}
The proof of this theorem is a simple exercise and we thus skip details. Let us just note that:\\[-3ex]
\begin{itemize}
\itemsep0pt
\item The case $s=+1$ corresponds to $M$ having two real eigenvalues equal to $\k\pm\l$.
\item The case $s=-1$ corresponds to $M$ having two complex-conjugate eigenvalues, $\k\pm\im\l$.
\item The case $s=0$ corresponds to a single eigenvalue $\kappa$. In this case, the parameter $\lambda$ (if nonzero) can be removed by a change of basis and eq.~\eqref{theorem} is then exactly the usual Jordan form of the matrix $M$.\\[-3ex]
\end{itemize}


\section{Searching for Wess-Zumino terms}
\label{app:WZterms}

While the invariant part of the effective Lagrangian can be constructed straightforwardly using the MC form and tensor methods, the search for physically interesting theories featuring enhanced soft limits cannot be concluded before we classify possible quasi-invariant contributions to the effective Lagrangian, that is, terms invariant up to a gradient, commonly denoted as the WZ terms. We use Witten's construction, where the WZ terms are obtained as invariant Lagrangians in a spacetime of dimension one higher than the actual physical spacetime. In four spacetime dimensions, this amounts to classifying all invariant 5-forms that belong to the Lie algebra cohomology of the symmetry. We follow ref.~\cite{Goon:2012dy}, where Witten's approach was used to obtain the Galileon Lagrangians as WZ terms.


\subsection{Doubly-enhanced soft limit: spin-zero case}

We want to construct invariant closed 5-forms out of the components of the MC form~\eqref{MCspin0}. To that end, we first introduce the set of linearly independent, Lorentz-invariant 4-forms $e^i$, defined by
\begin{align}
\notag
(e^1,e^2,e^3,e^4,e^5)&\equiv\eps_{\k\l\m\n}(\om_P^\k\wedge\om_P^\l\wedge\om_P^\m\wedge\om_P^\n,\om_K^\k\wedge\om_P^\l\wedge\om_P^\m\wedge\om_P^\n,\om_K^\k\wedge\om_K^\l\wedge\om_P^\m\wedge\om_P^\n,\\
\label{ei}
&\phantom{{}\equiv{}}\om_K^\k\wedge\om_K^\l\wedge\om_K^\m\wedge\om_P^\n,\om_K^\k\wedge\om_K^\l\wedge\om_K^\m\wedge\om_K^\n)\\
\notag
&\equiv\eps\cdot(\om_P^4,\om_K\wedge\om_P^3,\om_K^2\wedge\om_P^2,\om_K^3\wedge\om_P,\om_K^4).
\end{align}
Note that in case of the Galileon algebra, where $\om^\m_P=\dd x^\m$, $\om^\m_K=\dd\x^\m$ and $\om_Q=\dd\t-\x\cdot\dd x$, all the 5-forms $\om_5^i\equiv\om_Q\wedge e^i$ are trivially closed, and give rise to the five different Galileon terms in four spacetime dimensions~\cite{Goon:2012dy}. We will denote these compactly as
\begin{equation}
(\gal_1,\gal_2,\gal_3,\gal_4,\gal_5)\equiv(\dd\t-\x\cdot\dd x)\wedge(\dd x^4,\dd\x\wedge\dd x^3,\dd\x^2\wedge\dd x^2,\dd\x^3\wedge\dd x,\dd\x^4)\cdot\eps.
\label{galileon5form}
\end{equation}
With the presence of the additional generator $X$, the candidate 5-forms are forced by the assumed Lorentz invariance to be linear combinations of $\om_Q\wedge e^i$ and $\om_X\wedge e^i$.

To move on, we need to know the exterior derivatives of all the components of the MC form. Let us write the MC form generally as $\om=\om^iT_i$, where $T_i$ is the set of generators of the symmetry group. These components satisfy the MC structure equation
\begin{equation}
\dd\om^i=\frac12f^i_{jk}\om^j\wedge\om^k,
\label{MCeq}
\end{equation}
where $f^i_{jk}$ are the structure constants of the symmetry group, defined by $[T_i,T_j]=\im f^k_{ij}T_k$. We then infer that
\begin{equation}
\begin{split}
\dd\begin{pmatrix}
\om_P^\mu\\
\om_K^\mu\\
\end{pmatrix}&=    
\begin{pmatrix}
\k &  s\\
 1  &  \k\\
\end{pmatrix}
\begin{pmatrix}
\om_P^\mu\wedge\om_X\\
\om_K^\mu\wedge\om_X\\
\end{pmatrix},\\
\dd\om_Q&=\om_P^\m\wedge\om_{K\m}+2\k\om_Q\wedge\om_X,\\
\dd\om_X&=0.
\end{split}
\end{equation}
As a consequence, we find that
\begin{equation}
\dd e^i=-\om_X\wedge M^i_{\phantom ij}e^j,\qquad
\text{where}\qquad
M^i_{\phantom ij}\equiv\begin{pmatrix}
4\k & 4s & 0 & 0 & 0\\
1 & 4\k & 3s & 0 & 0\\
0 & 2 & 4\k & 2s & 0\\
0 & 0 & 3 & 4\k & s\\
0 & 0 & 0 & 4 & 4\k
\end{pmatrix}.
\label{dei}
\end{equation}
Note that all the 5-forms $\om_X\wedge e^i$ are trivially closed. On the other hand, once written in terms of the physical field $\t$, these lead to Lagrangian densities containing two derivatives per field, just like the leading invariant part of the Lagrangian. We therefore focus on the 5-forms $\om_Q\wedge e^i$, which should by construction contain $2n-2$ derivatives for $n$ factors of $\t$, and should therefore, if present, dominate the low-energy physics.

As a consequence of the linear independence of the 4-forms $e^i$ and of the fact that
\begin{equation}
\dd(\om_Q\wedge c_ie^i)=2\k\om_Q\wedge\om_X\wedge(c_ie^i)+(c_iM^i_{\phantom ij})\om_Q\wedge\om_X\wedge e^j,
\end{equation}
closed invariant 5-forms of the type $\om_5=\om_Q\wedge c_ie^i$ are in a one-to-one correspondence with the left eigenvectors of the matrix $M+2\k\mathbbm1$ with zero eigenvalue. This matrix has exactly one such eigenvector for any of the three allowed values of $s$ and $\k=0$. For $s\neq0$, it also has such an eigenvector for $\k=\pm\sqrt s/3$ and $\k=\pm2\sqrt s/3$. Since the parameter $\kappa$ determines the commutation relations of the real Lie algebra of the symmetry generators, it must itself be real; these extra solutions therefore only exists for $s=1$.

Let us first focus on the solutions for $s\neq0$. In this case, the unique closed 5-form $\om_5(s,\k)$ can be written in the neat form
\begin{equation}
\om_5(s,\k)=\om_Q\wedge\left(\om_K+\frac{\om_P}{\sqrt s}\right)^{2-3\k}\wedge\left(\om_K-\frac{\om_P}{\sqrt s}\right)^{2+3\k}\cdot\eps.
\end{equation}
It turns out that the dependence on the auxiliary field $\phi$ drops from all the forms. Put together with the single solution for $s=0$, the full list of closed invariant 5-forms $\om_5(s,\k)$ for various combinations of $s$ and $\k$ reads, in terms of the Galileon 5-forms~\eqref{galileon5form},
\begin{equation}
\begin{split}
\om_5(1,\pm\tfrac23)&=\gal_1\mp4\gal_2+6\gal_3\mp4\gal_4+\gal_5,\\
\om_5(1,\pm\tfrac13)&=-\gal_1\pm2\gal_2\mp2\gal_4+\gal_5,\\
\om_5(\pm1,0)&=\gal_1\mp2\gal_3+\gal_5,\\
\om_5(0,0)&=\gal_1.
\end{split}
\end{equation}
We can see that regardless of the values of $s$ and $\k$, the WZ term always contains $\gal_1$, which upon integration over the extra dimension translates into the tadpole term, $\La=\t$. We conclude that extending the Galileon algebra by an additional scalar generator $X$ does not lead to any nontrivial physical theories describing an interacting massless scalar.


\subsection{Doubly enhanced soft limit: spin-two case}

We proceed in the same manner as in the spin-zero case, this time omitting some of the straightforward technical details. Our task is again to construct invariant closed 5-forms, and the building blocks we now have at hand are $\om^{\m\n}_J$, $\om^\m_P$, $\om^{\m\n}_S$, $\om^\m_K$ and $\om_Q$, see eq,~\eqref{tracelessMCform}. The components $\om^{\m\n}_J$ and $\om^{\m\n}_S$ are uninteresting for the same reason as $\om_X$ in the spin-zero case: once expressed in terms of $\t$, they lead to Lagrangian densities containing two derivatives per field, just like the leading strictly invariant part of the Lagrangian.

Lorentz invariance then leaves us with $\om^i_5=\om_Q\wedge e^i$ as the only option, where $e^i$ take the same form (but different values due to different $\om^\m_P$ and $\om^\m_K$) as in eq.~\eqref{ei}. The equivalent of eq.~\eqref{dei} now reads
\begin{equation}
\dd e^i=\eps_{\k\l\m\n}\om^{\k\a}_S\wedge\om_{K\a}\wedge
\begin{pmatrix}
8s & 0 & 0\\
0 & 6s & 0\\
-\frac43 & 0 & 4s\\
0 & -6 & 0\\
0 & 0 & -24
\end{pmatrix}
\begin{pmatrix}
\om^\l_P\wedge\om^\m_P\wedge\om^\n_P\\
\om^\l_K\wedge\om^\m_P\wedge\om^\n_P\\
\om^\l_K\wedge\om^\m_K\wedge\om^\n_P
\end{pmatrix},
\end{equation}
and we have $\dd(c_ie^i)=-\om_Q\wedge(c_i\dd e^i)$. The $5\times3$ matrix of coefficients in the above equation has rank three, and thus has two left eigenvectors with eigenvalue zero. Upon some manipulation, the two corresponding closed 5-forms can be written as
\begin{equation}
\begin{split}
\om^1_5&=\om_Q\wedge(\dd x+\sqrt s\,\dd\x)^4\cdot\eps=\gal_1+4\sqrt s\gal_2+6s\gal_3+4s^{3/2}\gal_4+s^2\gal_5,\\
\om^2_5&=\om_Q\wedge(\dd x-\sqrt s\,\dd\x)^4\cdot\eps=\gal_1-4\sqrt s\gal_2+6s\gal_3-4s^{3/2}\gal_4+s^2\gal_5.
\end{split}
\end{equation}
These two have a unique combination that does not include the tadpole term $\gal_1$,
\begin{equation}
\om^1_5-\om^2_5\propto\gal_2+s\gal_4.
\end{equation}
We conclude that for any $s\in\{-1,0,+1\}$, there is a unique WZ term that arises from extending the Galileon algebra by an additional traceless symmetric tensor of redundant generators. For $s=\pm1$, this exactly reproduces the special Galileon~\cite{Hinterbichler:2015pqa}. For $s=0$, it is a mere kinetic term for the physical NG field $\t$.


\subsection{Multiple NG bosons and a single redundant generator}
 
 We want to see if it is possible to construct WZ terms out of the MC form~\eqref{GalileonMC}. To that end, we first write down the corresponding set of MC structure equations,
 \begin{equation}
\begin{split}
\dd\om_P^\m&=-h_i\Omega^i\wedge\om_K^\m,\\
\dd\om_K^\m&=-i_i\Omega^i\wedge\om_K^\m,\\
\dd\om_{\tilde Q}&=\om_P^\m\wedge\om_{K\m}-i_i\Omega^i\wedge\om_{\tilde Q},\\
\dd\Omega^i&=\frac12\str^i_{jk}\Omega^j\wedge\Omega^k.
\end{split}
\end{equation}
Lorentz invariance requires the candidate 5-forms to be built out of the 4-forms $e^i$~\eqref{ei}, wedged into $\omega_{\tilde Q}$ or a particular component of $\Omega$. However, the presence of the term $i_iK_\m$ in the commutator $[K_\m,\tilde Q_i]$, and likewise of $i_i\tilde Q$ in $[\tilde Q,\tilde Q_i]$, implies that upon an unbroken symmetry transformation $e^{\im\eps^\a\tilde Q_\a}$, both $\om^\m_K$ and $\om_{\tilde Q}$ receive a factor $e^{\eps^\a i_\a}$. Invariance of the WZ 5-forms under the \emph{internal} symmetry then requires that $i_\a=0$. Taking now, for instance, the set of 5-forms $\om_{\tilde Q}\wedge e^k$, we find upon a short calculation that
\begin{equation}
\dd(\om_{\tilde Q}\wedge e^k)=-ki_i\Omega^i\wedge\om_{\tilde Q}\wedge e^k-(5-k)h_i\Omega^i\wedge\om_{\tilde Q}\wedge e^{k+1}.
\end{equation}
This does not lead to any closed 5-forms unless $i_i\Omega^i=h_i\Omega^i=0$. Given the already known constraints $i_\a=h_\a=0$, this in turn implies that whether or not some of the generators $\tilde Q_i$ are spontaneously broken, the coefficients $h_i$ and $i_i$ must be zero. The only possibility how to construct WZ terms for the Lie-algebraic structure~\eqref{generalized_DBI} is therefore to take a direct sum of the simplest Galileon algebra with an additional internal symmetry. Any interactions between the Galileon and the non-Galileon NG sectors must then occur via strictly invariant terms in the Lagrangian, where the enhanced soft limit of scattering amplitudes for the Galileon mode is realized trivially.\footnote{There is an exception to the above argument in that the closure of $\om_{\tilde Q}\wedge e^5$ only requires $i_i\Omega^i=0$. This does not lead to any physically interesting theories either, though, since $\om_{\tilde Q}\wedge e^5$ cannot give rise to a kinetic term for the Galileon field.}


\bibliographystyle{JHEP}
\bibliography{references}

\end{document}